\documentclass[journal]{IEEEtran}
\usepackage{cite}

\ifCLASSINFOpdf
\usepackage[pdftex]{graphicx}
\else
\fi
\usepackage{epstopdf}
\usepackage{amssymb,amsmath}
\usepackage{xcolor}
\newcounter{storeeqcounter1}
\newcounter{storeeqcounter2}
\newcounter{storeeqcounter3}
\newcounter{storeeqcounter4}
\newcounter{tempeqcounter}

\newcommand{\MeijerG}[7]{G \begin{smallmatrix} #1#2 \\ #3#4 \end{smallmatrix} 
\left( #7 \middle\vert
\begin{matrix} #5 \\ #6 \end{matrix} \right) } 
\begin{document}
\title{Linear/Non-Linear Energy Harvesting Models via Multi-Antenna Relay Cooperation in V2V Communications}
\author{
Semiha~Kosu, Mohammadreza~Babaei, Serdar~\"{O}zg\"{u}r~Ata,
L\"{u}tfiye~Durak-Ata, and~Halim~Yanikomeroglu}

\maketitle
\begin{abstract}
Vehicle-to-vehicle (V2V) communications is a part of next-generation wireless networks to create smart cities with the connectivity of intelligent transportation systems. Besides, green communications is considered in V2V communication systems for energy sustainability and carbon neutrality. 
In this scope, radio-frequency (RF) energy harvesting (EH) provides a battery-free energy source as a solution for the future of V2V communications. 
Herein, the employment of RF-EH in V2V communications is considered where the bit error probability (BEP) of a dual-hop decode-and-forward relaying system is obtained depending on the utilization of antennas at the relay. The multiple antenna power-constraint relay harvests its power by applying dedicated antenna (DA)/power splitting (PS) EH modes and linear (L)/nonlinear (NL) EH models. Moreover, the links between nodes are exposed to double-Rayleigh fading. Finally, the performance of different system parameters is compared using theoretical derivations of BEP. 
The results provide a comprehensive analysis of the proposed system considering PS/DA-EH modes and L/NL-EH models, as well as deterministic/uniformly distributed placement of nodes.
It is observed that PS-EH outperforms DA-EH assuming a placement of an equal number of antennas and distances.
Moreover, optimal performance of PS/DA-EH is achieved by allocating more power and increasing the number of antennas for EH, respectively.
\end{abstract}
\IEEEpeerreviewmaketitle

\section{Introduction}
\IEEEPARstart{V}{ehicular} communications is a novel concept for next-generation wireless networks enhancing our future safety and comfort with intelligent transportation systems (ITS) applications. Moreover, ITS mostly provides users a driver-free experience in smart cities with the evaluation of future generation communications as well as traffic management, such as lane change warnings, automated parking applications, and online navigation systems. Additionally, vehicular communication systems offer an opportunity to transform the transportation industry by delivering ultra-reliable and low-latency connectivity for 6G use cases, allowing vehicles to communicate with each other and the surrounding infrastructure in real-time.
The most common types of communications among vehicles and beyond are mainly categorized as vehicle-to-vehicle (V2V) \cite{V2V1,V2V2,V2V3} and vehicle-to-infrastructure (V2I)\cite{V2I1,V2I2,V2V3}.

Energy consumption has become a major concern in next-generation wireless networks to assure green communications. Herein, radio-frequency (RF) energy harvesting (EH) technology can support mobile and vehicular networks while also addressing energy challenges by utilizing existing system tools without additional cost \cite{surveyEHv2v,Nasir13,DH-EH-reza}. 
The energy constraint nodes harvest incoming signal energy by applying simultaneous wireless information and power transfer (SWIPT) or wireless powered transfer (WPT) methods \cite{Nasir13,dipak2022,WPT-Reza}.
Here, both power-splitting (PS) and time-switching (TS) modes are the two types of SWIPT EH systems \cite{Nasir13,DH-EH-reza}.
In PS-EH mode, a portion of the incoming signal at an antenna is dedicated for EH while the remainder portion is allocated for the information processing (IP).
Moreover, in the TS EH mode, a fraction of the overall transmission time interval is allocated for EH and the remainder transmission time for IP. 
Furthermore, some antennas may be dedicated only for EH in wireless-powered communication (WPC) \cite{WPT-Reza} and then transmit with the harvested power.
Moreover, the amount of harvested energy is described as linear (L) EH or non-linear (NL) EH \cite{Nasir13,dipak2022,DH-EH-reza,piece2wise,PieceWiseModel_MIMO,WPT-Reza,nonlinear_exponential,nonlinear_AlouniModel,nonlinear_polynomialModel}.
In the L-EH model, the amount of harvested energy is directly proportional to the input power of the EH receiver \cite{Nasir13,dipak2022,DH-EH-reza}. 
However, in the NL-EH model, the amount of the harvested energy is saturated to a pre-defined threshold power for the high input power while the amount of harvested powers at low input power are equal for both L-EH and NL-EH (L/NL-EH) models \cite{WPT-Reza,piece2wise,PieceWiseModel_MIMO,nonlinear_exponential,nonlinear_AlouniModel,nonlinear_polynomialModel}.
\section{Related Works} \label{raltdwrk}
In the literature, cascaded fading channels are considered as the most suitable channel model for studying the effects for V2V communications \cite{Matolak,V2VMat} and have been extensively studied in \cite{impairments,generalfading,interferingV2V,kappamu, doublekappamu,NOMA,OutdatedCSI,CogCV2V,fulldup}.

Herein, some studies consider a point-to-point system in V2V communications when the channel is subjected to cascade fading \cite{impairments,generalfading,interferingV2V,kappamu,doublekappamu,NOMA}.
Outage probability (OP) expressions considering in-phase/quadrature-phase imbalances for a single/multiple carrier system at the transmitter and/or receiver are derived and compared with ideal transmitter and receiver over $N^*$Nakagami-$m$ fading channels in \cite{impairments}.
In \cite{generalfading}, a characteristic function-based approach for the average bit error rate (BER) of the equal-gain-combining diversity technique with binary modulation scheme in the presence of generalized fading models including double Rayleigh and double Nakagami-$m$ fading channels has been studied. 
\textcolor{black}{Moreover, physical layer security (PLS) analysis employed in V2V scenarios over cascaded fading channels is presented in \cite{interferingV2V,kappamu}. 
In \cite{interferingV2V}, the average secrecy capacity (ASC) and secrecy outage probability (SOP) of the system are analyzed to show the impact of interfering vehicles on the PLS in a V2V network over double Rayleigh fading channels. The special case of uniformly distributed spatial locations of vehicle nodes is demonstrated with the density model of the Poisson point process.
In addition, double $\kappa-\mu$ shadowed fading channels are demonstrated to investigate the SOP and ASC at the receiver using friendly jammer in \cite{kappamu}.}
In \cite{doublekappamu}, the probability density function (pdf) of the product of independently and non-identically distributed double $\kappa-\mu$ fading channels are obtained in the closed-form and compared to experimental data to characterize device-to-device and V2V communications.
Apart from traditional multiple access technologies, downlink non-orthogonal multiple access techniques for V2V communications are analyzed in \cite{NOMA} with respect to OP, average BER, and ergodic capacity when the channels between users are exposed to double Nakagami-$m$ fading. Here, the source is equipped with multiple antennas to ensure diversity and increase the system performance of the considered system with transmit antenna selection methods.

Due to the high mobility of vehicles, the coverage area for vehicles may change. Therefore, relay vehicles can improve the coverage for the end-user. Several studies investigated the cooperative V2V systems where the channels are subjected to cascaded fading \cite{OutdatedCSI,CogCV2V,fulldup}.
Two best relay selection, namely, predefined threshold and channel quality-based schemes have been applied in \cite{OutdatedCSI} to characterize the outdated channel state information caused by fast, time-varying fading attenuation and delay. Moreover, the relay applies a decode-and-forward (DF) relaying protocol over double Nakagami-$m$ fading channels to study the OP and average BER while achieving diversity order.
In \cite{CogCV2V}, the secondary network is equipped with source and destination vehicles, where both nodes have a direct link and a dual-hop (DH) link, while the relay applies the amplify-and-forward (AF) relaying protocol when the eavesdropping vehicle overhears the secondary transmitter. Herein, while the primary user interferes with the secondary user in underlay mode, the lower bound SOP and effective diversity order analysis are presented when the V2V/V2I links are exposed to double Rayleigh/Rayleigh fading channels.
In \cite{fulldup}, the lower-bound OP of a full-duplex (FD) cooperative communication network in the presence of relay self-interference is investigated where the relay applies the AF relaying protocol over cascaded Nakagami-$m$ fading channels. 

On the other hand, since the RF-EH is a promising solution for the sustainability in V2V communications, in the literature, L/NL-EH models are investigated in \cite{Nasir13,dipak2022,WPT-Reza,UAVlNL,nonlinear_exponential,kurma2021}.
The L-EH model, which overestimates the realistic performance of the energy constraint system are studied in \cite{Nasir13,dipak2022}.
In \cite{dipak2022}, authors investigated the OP of a secondary receiver in an EH cognitive radio (CR) network applying a multiple FD relaying protocol in a spectrum-sharing environment with multiple users. The impact of self interference at the receive antenna of each FD relay, as well as all transmitting interference at all receive nodes, is being addressed.
The OP and ergodic capacity of a DH AF relaying system is studied herein where the relay node applies PS and TS EH mode to harvest energy and then transmit the received signal via its harvested power. The authors examined the system performance considering different system parameters and showed that in most cases PS-EH mode provides better system performance. The NL-EH model is employed in \cite{WPT-Reza,UAVlNL}.
The BER performance of an overlay CR system where the secondary user improves the primary user performance is investigated in \cite{WPT-Reza}. The analytical derivations are derived considering L/NL-EH models under mixed Rayleigh/Nakagami-$m$ fading channels. Here, secondary user transmits the signal with the power that is harvested from the incoming signals from both primary transmitter and secondary user's receiver.
A coverage performance analysis of an energy constraint UAV-assisted SWIPT system is considered in \cite{UAVlNL} where ground nodes harvest energy applying PS/TS-EH modes. Moreover, lower bound theoretical expressions of coverage performance are derived for both L/NL-EH models where considered NL-EH model is adopted from \cite{nonlinear_exponential}. The authors showed that PS-EH mode outperforms the TS EH mode for all cases. 
In \cite{kurma2021}, a new transmission protocol that takes active user performance and NL-EH model into consideration is presented. The closed-form expression for the OP considering Rayleigh distributed fading channels is developed. In addition, the authors highlighted the effect of the number of cooperative users and other parameters on the performance of the system.

The integration of RF-EH capabilities with the V2V environment enables green communications as part of future communication systems, as well as sustainability and energy efficiency. Therefore, V2V-EH systems are considered in \cite{SecEHCog,afDF2020,AEU2022,Nguyen2021}.
In \cite{SecEHCog}, assuming an underlay CR network to improve security in the secondary transmitter's data link, a FD transmission is considered, while the secondary user (SU) adopts a PS L-EH model to generate jamming signals to reduce ASC in eavesdropper over cascaded $\kappa-\mu$ fading channels.
Joint and separate optimization of PS factor for a SWIPT based L-EH model is investigated, where both DF/AF relaying protocol is considered in the vehicular network \cite{afDF2020}. Moreover, the system performance is obtained by calculating average capacity and OP.
The performance of L-EH underlay CR multi-hop single antenna relay network is investigated in \cite{AEU2022}. Here, a dedicated power beacon node broadcasts energy-bearing signals to power-constraint nodes. In addition, TS-EH mode is applied for energy and data transmission through intermediate nodes. Finally, closed-form theoretical expressions of approximate and asymptotic OP are derived for SU.
The ergodic capacity of the FD AF/DF relaying system considering L-EH in V2V communication is studied in \cite{Nguyen2021}. The results show that the ergodic capacity is lower for the FD L-EH V2V communication system than for stationary nodes. The accurate closed-form expressions of ergodic capacity for both AF/DF relaying protocols over cascade (double) Rayleigh fading are derived theoretically.

In this paper, we combine the vehicular environment capabilities with an EH multiple-antenna relay-based system, and investigate the BER of the considered system with dedicated antenna (DA) and PS-EH modes for the DF relaying system. 
In DA-EH mode, some antennas of the relay are allocated for IP while the remaining antennas are dedicated to EH. 
On the other hand, in PS-EH mode, all antennas of the relay vehicle are used to some extent for both IP and EH.
Moreover, the BER of both L/NL-EH models are analyzed for the proposed system. 
Note that, the L-EH model misrepresents the realistic system performance for high input power of the EH receiver. 
In addition, the NL-EH model which is studied in \cite{piece2wise,PieceWiseModel_MIMO} is assumed since other NL-EH models are not mathematically tractable \cite{nonlinear_exponential,nonlinear_AlouniModel,nonlinear_polynomialModel}.
Since all nodes are mobile, here the suitable channel characteristic is determined to be worse than the Rayleigh fading in the literature \cite{Matolak} and exposed to a cascaded fading channel. 
In other words, field measurements have shown that the double Rayleigh fading channel model is a good fit for V2V communication channels. In this model, the wireless channel fading gain is described as the product of two independent Rayleigh distributed random variables \cite{V2VMat}.
Furthermore, since the line-of-sight (LoS) is not always guaranteed in vehicular communications, we assume that all of the corresponding channels between links are subjected to double Rayleigh fading \cite{S_kosu2019, CogCV2V}. To the best of our knowledge, EH has not been extensively studied jointly for V2V networks. 
Generally, the papers in \cite{SecEHCog,afDF2020,AEU2022,Nguyen2021} investigate only the outage or ergodic performance of the V2V-EH systems and BER performance metric is not addressed in \cite{SecEHCog,afDF2020,AEU2022,Nguyen2021}.
Moreover, \cite{SecEHCog,afDF2020,AEU2022,Nguyen2021} assume only the L-EH model with a single antenna, where some nodes are assumed to be static. However, this assumption does not completely reflect the realistic behavior of these systems since vehicles are mobile nodes. Additionally, the obtained results under the assumption of fixed distance for V2V scenarios are not reasonable and mislead our perception of V2V system design. Herein, we adopt the mobility of all nodes and assume that the distances between links are randomly distributed. In this way, V2V system performance analysis yields a more realistic approach.
The main contributions of this paper are summarized as follows:
\begin{itemize}
	\item The effect of randomly distributed nodes on the system performance is investigated and compared to the special case of deterministic distances between nodes.
    \item Theoretical derivations of the BER performance considering L/NL-EH models are derived over the double Rayleigh fading channels.
	\item Novel closed-form expressions for the BER analysis are derived for PS/DA-EH modes and comparisons between the results are provided.
	\item The analytical closed-form derivations show consistency with performed computer simulations considering different system parameters.
\end{itemize}
The remainder of this paper is organized as follows. Section \ref{System Model} describes the system model. In Section \ref{BEPanalysis}, the BER expressions of the L/NL-EH models considering the number of antennas at relay are derived and the application of the DA and PS modes are studied when the distances in the V2V network are deterministic and uniformly distributed. Section \ref{simResults} introduces the numerical and Monte-Carlo simulation results. Finally, the conclusion is presented in Section \ref{conclude}.

\textit{Notations:} $\mathop{\mathbb{E}}[\cdot]$ represents the expectation operator and $|\cdot|$ corresponds to the absolute value. Furthermore, $F_X(x)$ and $f_X(x)$ denote the cumulative distribution function (cdf) and pdf of a random variable $X$. Moreover, list of abbreviation is provided in Table \ref{abbrivat}.
\begin{table}[t!]
	\centering
	\caption{List of Abbreviation}
	\label{abbrivat}
	\resizebox{0.9\columnwidth}{!}
{%
	\renewcommand{\arraystretch}{1.1}
		\begin{tabular}{|c|l|}
\hline
AF   &Amplify-and-forward          \\ \hline
ASC  &Average secrecy capacity     \\ \hline
AWGN &Additive white Gaussian noise\\ \hline
BER  &Bit error rate                                  \\ \hline
cdf  & Cumulative distribution function \\ \hline
CR   &Cognitive radio\\ \hline
D    &Destination\\ \hline
DA   &Dedicated-antenna\\ \hline
DF   &Decode-and-forward\\ \hline
DH   &Dual-hop\\ \hline
EH   &Energy harvesting\\ \hline
FD   &Full-duplex\\ \hline
IP   &Information processing\\ \hline
ITS  &Intelligent transportation systems\\ \hline
L    &Linear                             \\ \hline
LoS  &Line-of-sight                     \\ \hline
MRC  &Maximum-ratio-combinig             \\ \hline
NL   &Non-linear                        \\ \hline
OP   &Outage probability              \\ \hline
pdf  &Probability density function        \\ \hline
PLS  &Physical layer security         \\ \hline
PS   &Power-splitting\\ \hline
R    &Relay\\ \hline
RF   &Radio-frequency\\ \hline
S    &Source\\ \hline
SER  &Symbol error probability\\ \hline
SNR  &Signal-to-noise ratio\\ \hline
SOP  &Secrecy outage probability\\ \hline
SU   &Secondary user\\ \hline
SWIPT&
Simultaneous wireless information and power transfer
\\ \hline
TS   &Time-switching\\ \hline
UAV  &Unmanned aerial vehicle\\ \hline
V2V  &Vehicle-to-vehicle\\ \hline
WPC  &Wireless-powered communication\\ \hline
WPT  &Wireless-powered transfer\\\hline
\end{tabular}%
}
\end{table}
\section{System Model}\label{System Model}
%
\begin{figure}[t!]
	\centering
	\includegraphics[width=1\columnwidth]{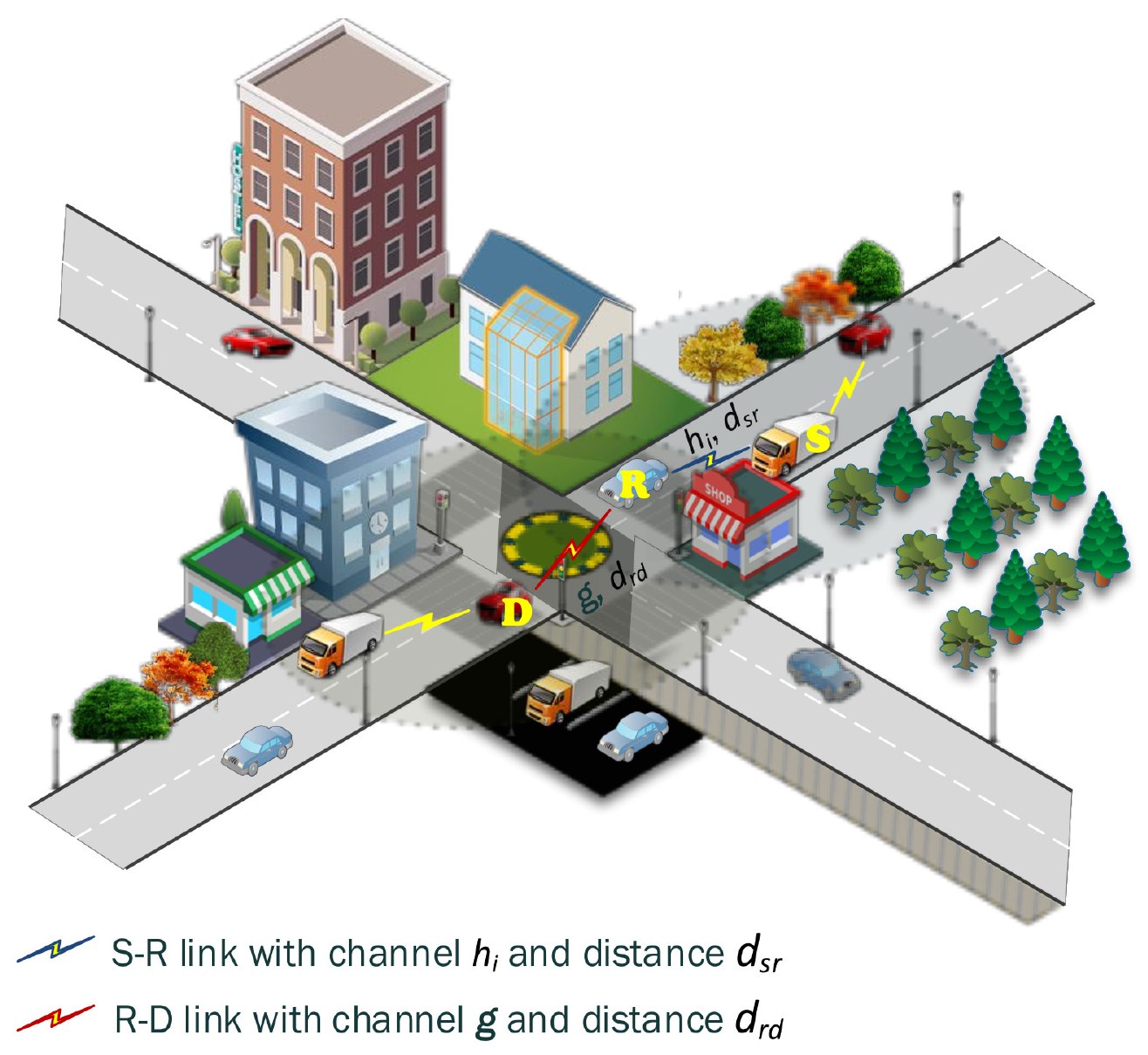}
	\caption{An energy constraint DH-DF relaying system model in which R is in the coverage zones of both S and D. The R harvests energy from S and uses its harvested power for data transmission.}
	\label{sysmoDeld}
\end{figure}

\begin{figure}[t!]
	\centering
	\includegraphics[width=1\columnwidth]{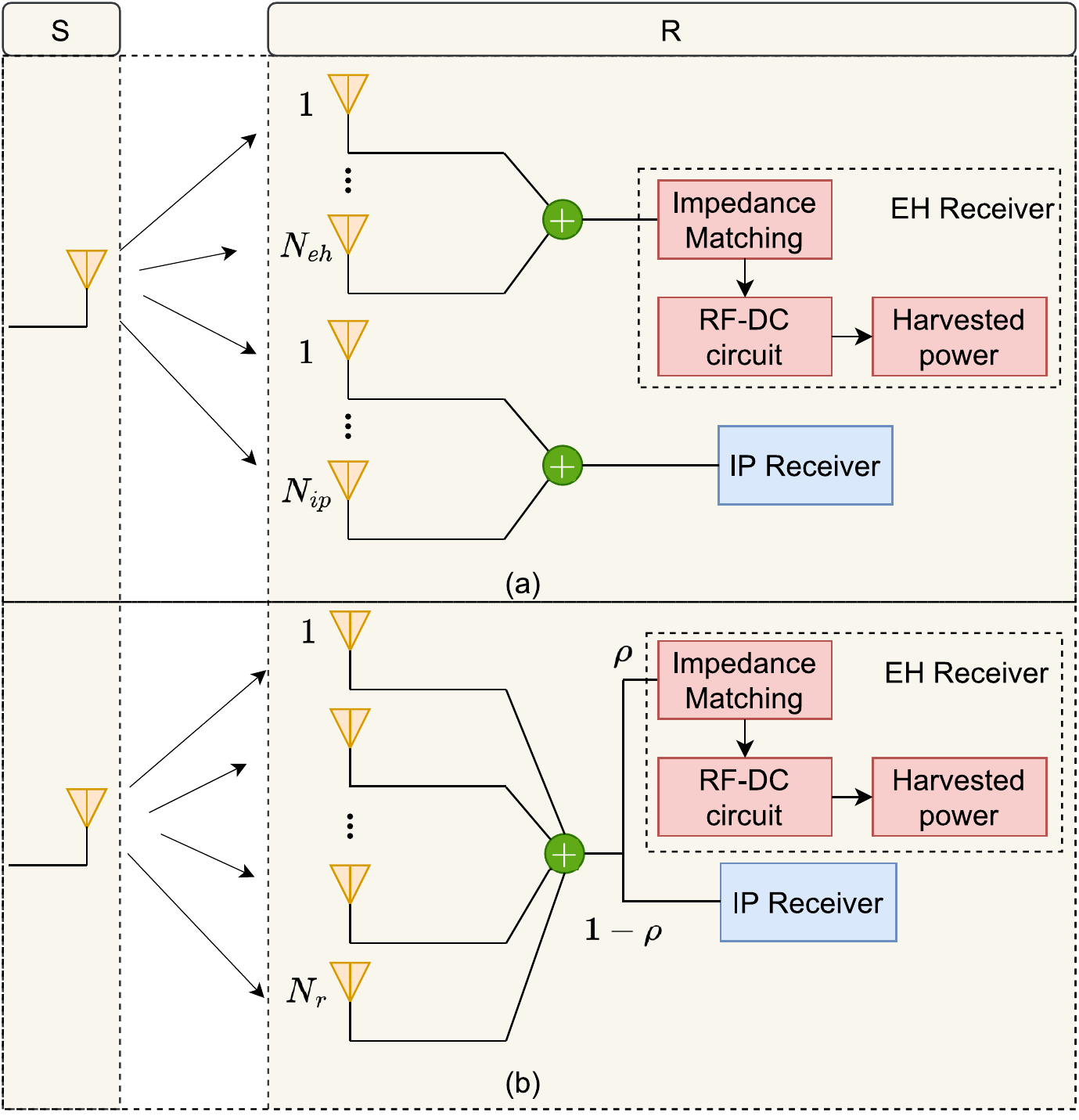}
	\caption{Description of the considered EH modes in the proposed V2V energy constraint system, (a) DA-EH mode where $N_{eh}$, $N_{ip}$ are the numbers of antennas dedicated for EH and IP, respectively, $N_r=N_{eh}+N_{ip}$ and $N_r$ is the number of receiving antennas at R, (b) PS-EH mode with $N_r$ antennas at R and EH factor $\rho$.}
		\label{DAbandPS}
\end{figure}

We consider a DH vehicular environment, in which all nodes are in motion as given in Fig. \ref{sysmoDeld}. 
Here, source (S) and destination (D) nodes are the master vehicles within their coverage areas, whereas the relay (R) is the closest nearby vehicle in the shared coverage of S and D. Furthermore, it is considered that communication between the S and D is critical since they are in charge of making centralized decisions for their clusters, and the neighboring R assists them in communicating without spending its own resources yet by harvesting energy.
Please note that, the proposed system model can be implemented in different real life scenarios such as blind spot, lane changing applications, and truck platooning \cite{surveyEHv2v}.
Moreover, S and D are equipped with one antenna while R is equipped with $N_r$ receive antennas and one transmit antenna applying maximum ratio combining (MRC) scheme.
Furthermore, the channels between all vehicles are considered as double Rayleigh distributed. The channels are independent at each block at a time and remain deterministic in one transmission interval of $T$.
Here, \textbf{$h_i$} and $g$ stand for the channels between S$\to$R and R$\to$D links, respectively, and $\Omega_h=\Omega_{h_i}, \Omega_g$ denote the channel gains where $i\in\{1,2,\cdots,N_r\}$.
Additionally, the distances between all vehicular nodes are uniformly distributed and compared with deterministic distances. 
Herein, $d_{sr}^{\omega}$ and $d_{rd}^{\omega}$ are the distances from S$\to$R, and R$\to$D links, respectively, where $\omega=\mathcal{D}$ or $\omega=\mathcal{U}$ denotes deterministic and uniformly distributed distances, respectively. 
For simplicity throughout the paper, it is assumed that the distances are denoted as $d_{sr}$ and $d_{rd}$.
Accordingly, the path-loss parameters are assigned as $L_{sr}=1/d_{sr}^v$ and $L_{rd}=1/d_{rd}^v$ where $v$ is the path-loss exponent.
For further consideration, the R forwards the information from the S to D using its own harvested energy and all the harvested energy is used to transmit from R. 

In all scenarios, EH is applied by considering two different modes: DA/PS-EH modes as in Fig. \ref{DAbandPS}(a), and Fig. \ref{DAbandPS}(b), respectively. In DA-EH mode, some antennas are allocated for EH while the remainders operate for IP illustrated in Fig. \ref{DAbandPS}(a).
On the other hand, all antennas at R simultaneously operate for both IP and EH in the PS-EH mode shown in Fig. \ref{DAbandPS}(b). The analysis of the EH modes has been done considering both L/NL-EH models. 
In the L-EH model, the input power and the harvested power are in direct proportion, thus an increase in input power results in an increase in the harvested power \cite{Nasir13,DH-EH-reza}. 
However, in the NL-EH model, for small amount of input power, the harvested power is in direct relation, while for high input power the harvested power is saturated to a threshold power level represented as $P_{th}$ \cite{piece2wise}. 
Without loss of generality, we denote the power spectral density of the additive white Gaussian noise (AWGN) as $N_0$ and, equal at all antennas of the receivers.
\subsection{DA-EH Mode}
Under the DA-EH mode, we assume that R is equipped with $N_r=N_{eh}+N_{ip}$ where $N_{eh}$ antennas operate for EH, while $N_{ip}$ antennas are reserved for IP.
In the first time slot, the received signal at R for IP is given as
\begin{align}
\textbf{y}_{sr}^{DA}=\sqrt{P_sL_{sr}}\textbf{h}_{ip}x+\textbf{n}_{ip}
\label{YsrAnten}
\end{align} 
where $\textbf{h}_{ip}=[h_1,h_2,\cdots,h_{N_{ip}}]$ and $\textbf{n}_{ip}=[n_1,n_2,\cdots,n_{N_{ip}}]$.
Moreover, $\textbf{n}_{i}$ stands for the sampled AWGN and $P_s$ represents the transmit power of S node.
Node R applies MRC algorithm to combine the transmitted signals $x$, then decodes the received signal and sends the estimated symbol $\hat{x}$ to node D.
Accordingly, the harvested power at node R is calculated as
\begin{align}
P_r^{DA}=\eta P_sL_{sr}\sum_{j=1}^{N_{eh}}|h_j|^2  
\label{Harvtspwreanten}
\end{align}
where $\eta$ is the energy conversion efficiency.
Furthermore, the received signal at D is expressed as 
\begin{align}
	{y}_{rd}^{DA}=\sqrt{P_r^{DA}L_{rd}}{g}\hat{x}+n_{rd}
	\label{yRD}
\end{align} 
where $n_{rd}$ is the sampled AWGN and $P_r^{DA}$ is calculated in \eqref{Harvtspwreanten}.
\subsection{PS-EH Mode}
%
For the PS mode, all $N_r$ antennas at node R simultaneously apply both IP and EH.
In the first time slot, the received signal at node R for the IP is given as 
\begin{align}
	\textbf{y}_{sr}^{PS}=\sqrt{(1-\rho)P_sL_{sr}}\textbf{h}x+\textbf{n}
	\label{YsrPS}
\end{align} 
where $\textbf{h}=[h_1,h_2,\cdots,h_{N_{r}}]$ and $\textbf{n}=[n_1,n_2,\cdots,n_{N_{r}}]$. 
Herein, $\rho$ represents the energy harvesting coefficient.
The harvested power at R is accordingly expressed as
\begin{align}
P_r^{PS}=\eta\rho P_sL_{sr}\sum_{j=1}^{N_{r}}|h_j|^2.
\label{HarvtspwrePS}
\end{align}
Similarly, after applying the MRC scheme at R when DF relaying is used for the transmission protocol between S and D, the received signal at D is given as
\begin{align}
	{y}_{rd}^{PS}=\sqrt{P_r^{PS}L_{rd}}{g}\hat{x}+n_{rd}.
	\label{yRDps}
\end{align} 
%

Considering \eqref{YsrAnten} and \eqref{YsrPS}, the received SNR at R for both PS/DA-EH modes is defined as
\begin{align}
\gamma _{sr}=\Theta Z
\label{snrSR}
  \end{align}
where
\begin{align}
	 \Theta\triangleq \, 
	 \begin{cases}
	 	\dfrac{P_s}{N_0}\sum_{i=1}^{N_{ip}}|h_i|^2, & {\text{DA-EH}} \\
	    \dfrac{(1-\rho)P_s}{N_0}\sum_{i=1}^{N_{r}}|h_i|^2, & {\text{PS-EH}}
	  \end{cases}
	  \label{tetaparamet}
  \end{align}
and $Z=L_{sr}$.
Here, under the assumption of uniformly distributed $d_{sr}$, $Z$ becomes a random variable.
Furthermore, considering \eqref{yRD}, and \eqref{yRDps}, the received SNR at node D for both DA/PS-EH modes are calculated as
\begin{align}
	\gamma_{rd}=XYZW
	\label{snrRD}
\end{align} 
where $Y=|g|^2/N_0$ and $W=L_{rd}$.
Please also note that, $W$ becomes a random variable since $d_{rd}$ is assumed to be uniformly distributed.
Here, $Y$ is the Double-Rayleigh fading with pdf given in \cite{karagihands} as
\begin{equation}
	f_Y(y)=\dfrac{2}{\bar{y}}\text{K}_0\left(2\sqrt{\dfrac{y}{\bar{y}}}\right)
	\label{fYy}
\end{equation} 
where $\bar{y}=\Omega_g/N_0$ and $\text{K}_v(.)$ is the modified Bessel function of the second kind with the $v-$th order. 
Please note that, unlike Rayleigh fading, which contains the exponential function, the double Rayleigh includes the Bessel function, which makes the theoretical derivations more complicated.
Moreover, we have 
\begin{equation}
	X\triangleq \, 
	\begin{cases}
		\eta P_s\sum_{j=1}^{N_{eh}}|h_j|^2, & {\text{DA-EH}} \\
		\eta\rho P_s\sum_{j=1}^{N_{r}}|h_j|^2, & {\text{PS-EH}}.
	\end{cases}
	\label{Xxparamet}
\end{equation}
Here, both $\Theta$ and $X$ in \eqref{tetaparamet} and \eqref{Xxparamet}, respectively, can be expressed as the sum of Gamma-Gamma distributions since MRC is applied at R. Then, the pdf is given as
\begin{align}
f_{\Delta}(\delta)=\psi_{\delta}\delta^{\alpha_{\delta}-1}\text{K}_{\xi_{\delta}}\left(2\sqrt{\beta_{\delta}\delta}\right)
\label{ftehxX}
\end{align} 
where $\Delta\in\{\Theta,X\}$ and $\delta\in\{\theta,x\}$ \cite{karagihands}.
Additionally, $\psi_{\delta}={2{\beta_{\delta}}^{\alpha_{\delta}}}/{\Gamma(\delta_{m_T})\Gamma(\delta_{k_T})}$, $\alpha_{\delta}=(\delta_{k_T}+\delta_{m_T})/2$, $\beta_{\delta}=\delta_{k_T}\delta_{m_T}/\bar{\delta}_T$, $\xi_{\delta}=\delta_{k_T}-\delta_{m_T}$ and $\epsilon=(-0.127- 0.95k-0.0058m)/(1+0.00124k+ 0.98m)$.
Please note that, for the double Rayleigh case we assume $k=m=1$.
The parameters of PS/DA-EH modes for S$\to$R, and R$\to$D links are provided in Table \ref{srlink} and Table \ref{rdlink}, respectively.  
\begin{table}[t!]
	\centering
	\caption{S$\to$R link parameters $\theta$}
	\label{srlink}
	\resizebox{0.9\columnwidth}{!}
{%
	\renewcommand{\arraystretch}{1.3}
		\begin{tabular}{|l|l|l|}
			\hline
			\multicolumn{1}{|c|}{S$\to$R link} & \multicolumn{1}{c|}{DA-EH}   & \multicolumn{1}{c|}{PS-EH}   \\ \hline
			$\bar{\theta}_T$ & $N_{ip}\bar{\theta}$   & $N_{r}\bar{\theta}$      \\ \hline
			$\bar{\theta}$   & $P_s\Omega_{h}/N_0$ & $(1-\rho) P_s\Omega_{h}/N_0$ \\ \hline
			$\theta_{k_T}$       & $N_{ip}k+(N_{ip}-1)\epsilon$        & $N_{r}k+(N_{r}-1)\epsilon$             \\ \hline
			$\theta_{m_T}$       & $N_{ip}m$                 & $N_{r}m$                      \\ \hline
		\end{tabular}%
}
\end{table}
\begin{table}[t!]
	\centering
	\caption{R$\to$D link parameters $X$}
	\label{rdlink}
	\resizebox{0.9\columnwidth}{!}{%
	\renewcommand{\arraystretch}{1.3}
		\begin{tabular}{|l|l|l|}
			\hline
			\multicolumn{1}{|c|}{R$\to$D link} & \multicolumn{1}{c|}{DA-EH}   & \multicolumn{1}{c|}{PS-EH}   \\ \hline
			$\bar{x}_T$ & $N_{eh}\bar{x}$           & $N_{r}\bar{x}$                \\ \hline
			$\bar{x}$   & $\eta P_s\Omega_{h}$ & $\eta\rho P_s\Omega_{h}$ \\ \hline
			$x_{k_T}$       & $N_{eh}k+(N_{eh}-1)\epsilon$        & $N_{r}k+(N_{r}-1)\epsilon$             \\ \hline
			$x_{m_T}$       & $N_{eh}m$                 & $N_{r}m$                      \\ \hline
		\end{tabular}%
	}
\end{table}
%
\section{BER Analysis} \label{BEPanalysis}
In digital communication systems, BER performance determines how reliable the communication is in the system.
The overall BER of the proposed DF relaying system is upper-bounded by 
\begin{equation}
	\mathbb{P}^{\mathcal{M}, \mathcal{N}}_{b}\leq
\dfrac{1-(1-\mathbb{P}_{sr}^\mathcal{N})(1-\mathbb{P}^{\mathcal{M}, \mathcal{N}}_{rd})}{\iota}
	\label{PeOverall}
\end{equation} 
where $\mathcal{M}\in\{\mathcal{L, NL}\}$, $\mathcal{L}$ and $\mathcal{NL}$ represents L/NL-EH models, respectively. 
Moreover, for $\mathcal{N}\in\{\mathcal{D, U}\}$, $\mathcal{D}$ and $\mathcal{U}$ represent the deterministic and uniformly distributed distances between the links S$\to$R and R$\to$D, respectively. 
Here, $\iota=\log_2M$, where $M$ corresponds to modulation order.
Moreover, $\mathbb{P}_{sr}^\mathcal{N}$, and $\mathbb{P}^{\mathcal{M}, \mathcal{N}}_{rd}$ in \eqref{PeOverall} represent symbol error rate (SER) of the links S$\to$R and R$\to$D, respectively.
\subsection{BER Analysis of S$\to$R Link Over Uniformly Distributed $d_{sr}^\mathcal{U}$}
The conditional SER of S$\to$R link considering \eqref{snrSR} is given as
\begin{align}
    \mathbb{P}_{sr}^{\mathcal{U}}(e|Z)=\int_{0}^{\infty}a\text{Q}(\sqrt{2b\Theta Z})f_\Theta(\theta)\text{d}\theta
    \label{PeconditnZint}
\end{align}
where $a$ and $b$ are the modulation specific parameters\cite{DH-EH-reza}, and $\text{Q}(.)$ is the Q-function \cite{Ryzhik43}. Using \cite[eq. 06.27.26.0006.01]{erfc} and \cite[eq. 9.31-1]{Abramowitzz}, substituting $f_{\Theta}(\theta)$ from \eqref{ftehxX} and applying \cite[eq. 14]{Adamchik}, \cite[eq. 9.31-1]{Abramowitzz}, and \cite[eq. 21]{Adamchik}, we have 
\begin{align}
	&\mathbb{P}_{sr}^{\mathcal{U}}(e|Z)=\nonumber\\&\tau_{\theta}
\MeijerG{3,}{3}{5,}{4}{1,1-\alpha_{\theta}-\dfrac{\xi_{\theta}}{2},1-\alpha_{\theta}+\dfrac{\xi_{\theta}}{2},-\alpha_{\theta},1}{0,0.5,-\alpha_{\theta},1}{\dfrac{bz}{\beta_{\theta}}}
\label{PeconditnZClosed}
\end{align}
where $\tau_{\theta}={a\psi_{\theta}}/({4\sqrt{\pi}{\beta_{\theta}}^{\alpha_{\theta}}})$.
Averaging over $Z$ in \eqref{PeconditnZClosed}, we obtain 
\begin{equation}
	\mathbb{P}_{sr}^{\mathcal{U}}(e)=\int_{g^{-v}}^{f^{-v}}\mathbb{P}_{sr}^{\mathcal{U}}(e|Z)f_Z(z)\text{d}z=\dfrac{\tau_{\theta}}{(g-f)v}(\Xi(f)-\Xi(g))
	\label{peuniformclosed}
\end{equation} 
where $\Xi(\kappa)$ is the function with $\kappa\in\{f,g\}$, and $d_{sr}^\mathcal{U}\sim\mathcal{U}(f,g)$.
Moreover, $\Xi(\kappa)$ is given in
\eqref{SRSEP_fix}
\addtocounter{equation}{1}%
\setcounter{storeeqcounter1}%
{\value{equation}}%
on top of Page~\pageref{SRSEP_fix}.
\begin{figure*}[!t]
\normalsize
\setcounter{tempeqcounter}{\value{equation}} 
\begin{IEEEeqnarray}{rCl}
\setcounter{equation}{\value{storeeqcounter1}} 
\Xi(\kappa)= \kappa
\MeijerG{3,}{4}{6,}{5}{1,1-\alpha_{\theta}-\xi_{\theta}/2,1-\alpha_{\theta}+\xi_{\theta}/2,1+1/v,-\alpha_{\theta},1}{0,0.5,-\alpha_{\theta},1/v,1}{\dfrac{b}{{\kappa}^v\beta_{\theta}}}.
\label{SRSEP_fix}
\end{IEEEeqnarray}
\setcounter{equation}{\value{tempeqcounter}} 
\hrulefill
\vspace*{4pt}
\end{figure*}
Here $\Xi(\kappa)$ is calculated from \cite[eq. 26]{Adamchik} where $f_Z(z)$ is obtained from \eqref{pdfZ} in Appendix \ref{appendixa}.
Based on the obtained expressions given in \eqref{peuniformclosed} and \eqref{SRSEP_fix} considering uniformly distributed $d_{sr}^\mathcal{U}$, the SER for the S$\to$R link for both PS/DA-EH modes depends on the $N_r$, and channel characteristic parameters, as well as $P_s$.
\subsection{BER Analysis of S$\to$R Link Over Deterministic $d_{sr}^\mathcal{D}$}
Please note that, for a deterministic distance, the $z$ parameter in \eqref{PeconditnZClosed} is deterministic. 
In other words, we replace $Z=L_{sr}$ in \eqref{PeconditnZClosed} since $Z$ in a fixed value. Hence, the SER of the S$\to$R link is calculated as
\begin{align}
\mathbb{P}_{sr}^{\mathcal{D}}(e)=\mathbb{P}_{sr}^{\mathcal{U}}(e|Z=L_{sr}).
\label{Peconditnfixedsr}
\end{align}
Here, we obtained the closed-form expression for the SER of S$\to$R for both PS/DA-EH modes. It is perceived that the system performance is affected by $N_r$, channel parameters, and $P_s$.

\subsection{BER Analysis of R$\to$D Link Over Uniformly Distributed $d_{rd}^\mathcal{U}$}
Considering the received SNR at node D given in \eqref{snrRD}, the conditional SER of R$\to$D link is written as 
\begin{align}
    \mathbb{P}_{rd}^\mathcal{U}(e|X,W,Z)=\int_{0}^{\infty}a\text{Q}(\sqrt{2bxywz})f_Y(y)\text{d}y.
    \label{rdUcondixwz}
\end{align}
Using \cite[eq. 06.27.26.0006.01]{erfc}, and \cite[eq. 9.31-1]{Abramowitzz}, substituting $f_{Y}(y)$ from \eqref{fYy} and applying \cite[eq. 14]{Adamchik}, \cite[eq. 9.31-1]{Abramowitzz}, and \cite[eq. 21]{Adamchik}, we obtain 
\begin{equation}
	\mathbb{P}_{rd}^{\mathcal{U}}(e|X,W,Z)=\dfrac{a}{2\sqrt{\pi}}\MeijerG{3,}{3}{5,}{4}{1,0,0,-1,1}{0,0.5,-1,1}{b\bar{y}xzw}
	\label{peXclosedform}
\end{equation} 
where $\MeijerG{m,}{n}{p,}{q}{.}{.}{.}$ is the Meijer's-G function \cite{Ryzhik43}. 
For simplicity, we assume $U=XZ$ where the pdf of $f_U(u)$ is derived in Appendix \ref{appendixbfUu}.
Please note that, $U$ stands for the harvested power at R considering both \eqref{Harvtspwreanten} and \eqref{HarvtspwrePS} for DA/PS-EH modes, respectively.
Further, averaging over $w$, we have
\begin{align}
	\mathbb{P}_{rd}^{\mathcal{U}}(e|U)=\int_{p^{-v}}^{r^{-v}}\mathbb{P}_{rd}^{\mathcal{U}}(e|U,W)f_W(w)\text{d}w=\varrho(r)-\varrho(p)
	\label{rdcondxU}
\end{align} 
where $f_W(w)$ is calculated in Appendix \ref{appendixa}.
Moreover, employing \cite[eq. 26]{Adamchik}, $\varrho(.)$ in \eqref{rdcondxU} can be written as 
\begin{align}
\varrho(\mu)&=\dfrac{a\mu}{2\sqrt{\pi}(p-r)v}\nonumber\\&\times\MeijerG{3,}{4}{6,}{5}{1,0,0,1+1/v,-1,1}{0,0.5,-1,1/v,1}{b\bar{y}u/{\mu}^v}
\end{align} 
where $\varrho(\mu)$ is the function with $\mu\in\{r,p\}$ and $d_{rd}^\mathcal{U}\sim\mathcal{U}(r,p)$.
Finally, the conditional SER for the uniformly distributed $d_{rd}^\mathcal{U}$ of R$\to$D link is derived in \eqref{rdcondxU} for both DA/PS-EH modes.
%
\subsubsection{L-EH model with Uniformly Distributed $d_{rd}^{\mathcal{U}}$}  
Considering the L-EH model and averaging over $U$ in \eqref{rdcondxU}, we have 
\begin{align}
	\mathbb{P}_{rd}^{{\mathcal{L}},{\mathcal{U}}}(e)&=\int_{0}^{\infty}\mathbb{P}_{rd}^{\mathcal{U}}(e|U)f_U(u)\text{d}u=\dfrac{\psi_{x}}{2v(g-f)}
	\left(\varphi(r,f)\right.\nonumber\\&\left.
	-\varphi(r,g)-\varphi(p,f)+\varphi(p,g)\right)
	\label{rdLUintegral}
\end{align} 
where $\varphi(\mu,\kappa)$ is the function with $\mu\in\{r,p\}$, and $\kappa\in\{f,g\}$.
Substituting, $f_U(u)$ and $\mathbb{P}_{rd}^{\mathcal{U}}(e|U)$ from Appendix \ref{appendixbfUu}, and \eqref{rdcondxU}, respectively,
\eqref{rdLUintegral} is re-written as \eqref{mmu2n2int}
\addtocounter{equation}{1}%
\setcounter{storeeqcounter2}%
{\value{equation}}%
on top of Page~\pageref{mmu2n2int}.
\begin{figure*}[!t]
\normalsize
\setcounter{tempeqcounter}{\value{equation}} 
\begin{IEEEeqnarray}{rCl}
\setcounter{equation}{\value{storeeqcounter2}} 
\varphi(\mu,\kappa)&=&\dfrac{a\mu{\kappa}^{1+v\alpha_x}}{2v(p-r)\sqrt{\pi}}
\int_0^{\infty}u^{\alpha_x-1}
\MeijerG{3,}{1}{2,}{4}{1,1-\alpha_x-1/v}{0.5\xi_x,-0.5\xi_x,-\alpha_x-1/v,1}{\beta_x{\kappa}^v {u}}
\nonumber\\&\times&
\MeijerG{3,}{4}{6,}{5}{1,0,0,1+1/v,-1,1}{0,0.5,-1,1/v,1}{\dfrac{b\bar{y}u}{{\mu}^v}}\text{d}u
\label{mmu2n2int}
\end{IEEEeqnarray}
\setcounter{equation}{\value{tempeqcounter}} 
\hrulefill
\vspace*{4pt}
\end{figure*}
Using \cite[eq. 14]{Adamchik}, \cite[eq. 9.31-1]{Abramowitzz} and \cite[eq. 21]{Adamchik} in \eqref{mmu2n2int}, the closed-form derivation of \eqref{rdLUintegral} is provided in \eqref{mmu2n2closed} given  
\addtocounter{equation}{1}%
\setcounter{storeeqcounter3}%
{\value{equation}}%
on top of
Page~\pageref{mmu2n2closed}.
\begin{figure*}[!t]
\normalsize
\setcounter{tempeqcounter}{\value{equation}} 
\begin{IEEEeqnarray}{rCl}
\setcounter{equation}{\value{storeeqcounter3}} 
\varphi(\mu,\kappa)&&=
\dfrac{a\mu{\kappa}^{1+v\alpha_x}}{2v(p-r)\sqrt{\pi}}
{({\mu}^v/b\bar{y})}^{\alpha_{x}}\nonumber\\&&\times\MeijerG{7,}{4}{7,}{10}
{1,1-\alpha_x,0.5-\alpha_x,2-\alpha_x,1-\alpha_x-1/v,-\alpha_x,1-\alpha_x-1/v}
{0.5\xi,-0.5\xi,-\alpha_x-1/v,-\alpha_x,1-\alpha_x,1-\alpha_x,-\alpha_x-1/v,2-\alpha_x,-\alpha_x,1}
{\dfrac{\beta_x{\kappa}^v{\mu}^v}{b\bar{y}}}
\label{mmu2n2closed}
\end{IEEEeqnarray}
\setcounter{equation}{\value{tempeqcounter}} 
\hrulefill
\vspace*{4pt}
\end{figure*}
Hence, the theoretical derivations of R$\to$D link considering the uniformly distributed $d_{rd}^\mathcal{U}$ is obtained. 
Finally, the upper bound BER of the uniformly distributed distance under the L-EH model is calculated by substituting \eqref{peuniformclosed}, and \eqref{rdLUintegral} in \eqref{PeOverall}. 
The obtained general theoretical expressions are valid for both PS/DA-EH modes. Moreover, the desired parameters are provided in Tables \ref{srlink} and \ref{rdlink}.
\subsubsection{NL-EH model with Uniformly Distributed $d_{rd}^{\mathcal{U}}$}
For the NL-EH model, the harvested power is saturated to $P_{th}$ for input powers which are higher than the predefined threshold value of $P_{th}$ \cite{piece2wise}. 
Hence, unlike the L-EH model, the SER of the link R$\to$D is calculated as a sum of two terms. One term provides the SER where the harvested power is below the threshold value, and the other term provides the SER where the harvested power is above the threshold power. Hence, we have
\begin{align}
	\mathbb{P}_{rd}^{{\mathcal{NL}},{\mathcal{U}}}(e)=\mathbb{J}_1+\mathbb{P}_{rd}^{\mathcal{U}}(e|U=P_{th})\left(1-\mathbb{J}_2\right).
	\label{PrdNLu}
\end{align}
Please note that the term $\mathbb{J}_1$ calculates the SER for a region where input power at the EH circuit is smaller than the threshold. This region characterizes L-EH model while the second term $\mathbb{P}_{rd}^{\mathcal{U}}(e|U=P_{th})\left(1-\mathbb{J}_2\right)$ represents the SER calculation for higher input powers at EH circuit. Furthermore, $\mathbb{P}_{rd}^{\mathcal{U}}(e|U=P_{th})$ is derived by using \eqref{rdcondxU} while taking $U=P_{th}$. Moreover, $\mathbb{J}_2$ is calculated as 
\begin{align}
\mathbb{J}_2&=\int_{0}^{P_{th}}f_U(u)\text{d}u
=\dfrac{\psi_{x}}{2v(g-f)}(\sigma(f)-\sigma(g))
\label{J2sigmas}
\end{align}  
where $f_U(u)$ is obtained in \eqref{fUufirst} from Appendix \ref{appendixbfUu} where $U=XZ$ and $d_{sr}^\mathcal{U}\sim\mathcal{U}(f,g)$.
Furthermore, considering \cite[eq. 26]{Adamchik}, $\sigma(\kappa)$ in \eqref{J2sigmas} is calculated as 
\begin{align}
	\sigma(\kappa)&={P_{th}}^{\alpha_x}{\kappa}^{1+v\alpha_x}
	\nonumber\\&\times
	\MeijerG{3,}{2}{3,}{5}
	{1,1-\alpha_x,1-\alpha_x-1/v}
	{\frac{\xi_x}{2},-\frac{\xi_x}{2},-\alpha_x-1/v,-\alpha_x,1}
	{\beta_x P_{th}{\kappa}^v}
	\label{CDFx}
\end{align}  
where $\kappa\in\{f,g\}$.
The term $\mathbb{J}_1$ in \eqref{PrdNLu} is similar to \eqref{rdLUintegral} apart from the integral boundaries. 
This is due to the fact that the region below $P_{th}$ characterize as the L-EH model. Hence, we have
\begin{align}
	\mathbb{J}_1=\int_{0}^{P_{th}}\mathbb{P}_{rd}^{\mathcal{U}}(e|U)f_U(u)\text{d}u.
	\label{J1cheyb_ilk}
\end{align}
However, there is no closed-form solution over the interval $(0, P_{th})$. Therefore, Gaussian-Chebyshev quadrature approximation is considered to be an effective approach to solve the integral given in \eqref{J1cheyb_ilk}.
In this approach, any given function $f(x)$ over the integral $(a,b)$ can be defined as $\int_{a}^{b}f(x) \text{d}x \approx \frac{b-a}{2} \sum_{i=1}^{n} w_i \sqrt{1-y_i^2} f(x_i)$, where $x_i=\frac{b-a}{2} y_i + \frac{b+a}{2}$, $y_i=\cos \frac{2i-1}{2n}\pi$, and $w_i=\frac{\pi}{n}$ \cite{Chebishev}. 
Moreover, applying Gaussian-Chebyshev approximation in \eqref{J1cheyb_ilk}, we obtain
\begin{align}
	\mathbb{J}_1=\dfrac{P_{th}}{2}\sum_{i=1}^{\chi}\dfrac{\pi}{\chi}\sqrt{1-y_i^i}f(x_i)
	\label{J1cheyb}
\end{align}
where $x_i=0.5P_{th}y_i+0.5P_{th}$, $y_i=\cos((2i-1)\pi/2\chi)$.
Here, $\chi$ is the parameter which determines a trade-off between complexity and accuracy. 
Subsequently, by substituting $\mathbb{P}_{rd}^{\mathcal{U}}(e|U=P_{th})$ from \eqref{rdcondxU}, \eqref{J2sigmas} with \eqref{CDFx}, and \eqref{J1cheyb}, the SER of the NL-EH model of the link R$\to$D with uniformly distributed $d_{rd}^{\mathcal{U}}$
 is obtained by using \eqref{PrdNLu}.

Finally, the upper-bound BER of the proposed system considering uniformly distributed distance under the NL-EH model is calculated by applying \eqref{peuniformclosed}, and \eqref{PrdNLu} in \eqref{PeOverall}.
Hence, the obtained theoretical expressions provide an insight through the NL-EH model into the considered system performance.
\subsection{BER Analysis of R$\to$D Link Over Deterministic $d_{rd}^\mathcal{D}$}
The conditional SER of R$\to$D link for the deterministic distance $d_{rd}^\mathcal{D}$ using \eqref{peXclosedform} and considering $Z=L_{sr}$, and $W=L_{rd}$ is expressed as
\begin{align}
	\mathbb{P}_{rd}^{\mathcal{D}}(e|\Phi)=\dfrac{a}{2\sqrt{\pi}}\MeijerG{3,}{3}{5,}{4}{1,0,0,-1,1}{0,0.5,-1,1}{b\bar{y}L_{rd}\Phi}
	\label{prdFconditXint}
\end{align} 
where $\Phi=L_{sr}X$.
Here, $\Phi$ stands for the harvested power at R considering the deterministic distance $d_{rd}^\mathcal{D}$ for both \eqref{Harvtspwreanten} and \eqref{HarvtspwrePS} and DA/PS-EH modes, respectively. Moreover, $f_\Phi(\phi)=f_X(\phi/L_{sr})/L_{sr}$ is obtained applying \eqref{ftehxX} as
\begin{align}
	f_{\Phi}(\phi)=
	\psi_{\phi}
	\phi^{\alpha_{x}-1}
	\text{K}_{\xi_{x}}(2\sqrt{\beta_{\phi}\phi})
	\label{fphifi}
\end{align} 
where $\psi_{\phi}=\psi_{x}/({L_{sr}})^{\alpha_{x}}$, and $\beta_{\phi}=\beta_{x}/L_{sr}$.
Please note that, unlike uniformly distributed $d_{rd}^\mathcal{U}$ where $U=XZ$, for the case of $d_{rd}^\mathcal{D}$, $Z$ is deterministic and the harvested power is defined as $\Phi=L_{sr}X$.
\subsubsection{L-EH model with Deterministic Distance $d_{rd}^\mathcal{D}$}
For the L-EH model, the SER of the link R$\to$D is calculated as 
\begin{equation}
	\mathbb{P}_{rd}^{\mathcal{L},\mathcal{D}}(e)=\int_{0}^{\infty}\mathbb{P}_{rd}^{\mathcal{D}}(e|\Phi)f_{\Phi}(\phi)\text{d}\phi.
	\label{Prd}
\end{equation} 
Please note that different from NL-EH model, the SER is defined by averaging over integral boundary $(0,\infty)$.
Substituting \eqref{prdFconditXint}, and \eqref{fphifi} in \eqref{Prd} and using \cite[eq.14]{Adamchik}, \cite[eq.9.31-1]{Abramowitzz} and \cite[eq.21]{Adamchik}, we have \eqref{PrdLfclosedForm} which is given 
\addtocounter{equation}{1}%
\setcounter{storeeqcounter4}%
{\value{equation}}%
on top of Page~\pageref{PrdLfclosedForm}.
\begin{figure*}[!t]
	\normalsize
	\setcounter{tempeqcounter}{\value{equation}} 
	\begin{IEEEeqnarray}{rCl}
		\setcounter{equation}{\value{storeeqcounter4}} 
	\mathbb{P}_{rd}^{\mathcal{L},\mathcal{D}}(e)&=&\dfrac{a\psi_\phi}{4\sqrt{\pi}}
	\MeijerG{5,}{4}{5,}{8}{1,1-\alpha_x,0.5-\alpha_x,2-\alpha_x,-\alpha_x}{\frac{\xi_x}{2},\frac{-\xi_x}{2},-\alpha_x,1-\alpha_x,1-\alpha_x,2-\alpha_x,-\alpha_x,1}{\dfrac{\beta_\phi}{bL_{rd}\bar{y}}}/\left(bL_{rd}\bar{y}\right)^{\alpha_x}
\label{PrdLfclosedForm}
	\end{IEEEeqnarray}
	\setcounter{equation}{\value{tempeqcounter}} 
	\hrulefill
	\vspace*{4pt}
\end{figure*}
Finally, substituting \eqref{Peconditnfixedsr} and \eqref{PrdLfclosedForm} in \eqref{PeOverall}, the overall BER of the considered system is obtained. 
Based on the closed-form theoretical derivations, BER depends on the $N_r$, and, channel parameters, as well as $P_s$.
\subsubsection{NL-EH model with Deterministic Distance $d_{rd}^\mathcal{D}$}
For the NL-EH model, SER of the link R$\to$D is given as
\begin{align}
	\mathbb{P}_{rd}^{\mathcal{NL},\mathcal{D}}(e)=\mathbb{I}_1+\mathbb{P}_{rd}^{\mathcal{D}}(e|\Phi=P_{th})+(1-\mathbb{I}_2)
	\label{prdNLfintfirst}
\end{align}
where $\mathbb{P}_{rd}^{\mathcal{D}}(e|\Phi=P_{th})$ is derived using \eqref{prdFconditXint} by taking $\Phi=P_{th}$.
Moreover, in \eqref{prdNLfintfirst}, after substituting $f_\Phi(\phi)$, and $\mathbb{P}_{rd}^{\mathcal{D}}(e|\Phi)$ from \eqref{fphifi}, and \eqref{prdFconditXint}, and considering \cite[eq. 03.04.06.0002.01]{erfc}, we have 
\begin{align}
	\mathbb{I}_1&=\int_{0}^{P_{th}}\mathbb{P}_{rd}^{\mathcal{D}}(e|\Phi)f_\Phi(\phi)\text{d}\phi\nonumber\\&=\dfrac{a\psi_\phi\sqrt{\pi}\csc(\pi\xi_x)}{4}(\Psi(-\xi)-\Psi(\xi))
	\label{I1closed}
\end{align}
where  
\begin{align}
	\Psi(\zeta)&=\sum_{j=0}^{\chi}\dfrac{{{\beta_\phi}}^{j+0.5\zeta}}{\Gamma(j+\zeta+1)j!}\int_{0}^{P_{th}}x^{\alpha_x+j+\zeta/2-1}\nonumber\\&
	\times\MeijerG{3,}{3}{5,}{4}{1,0,0,-1,1}{0,0.5,-1,1}{bxL_{rd}\bar{y}}\text{d}x,
	\label{I11NLfint}
\end{align}
and $\zeta\in\{-\xi_x,\xi_x\}$. 
Using \cite[eq. 26]{Adamchik}, \eqref{I11NLfint} is calculated as 
\begin{align}
	\Psi(\zeta)&=\sum_{j=0}^{\chi}\dfrac{{\beta_\phi}^{j+0.5\zeta}{P_{th}}^{\alpha_x+j+\zeta/2}}{\Gamma(j+\zeta+1)j!}\nonumber\\&
	\times\MeijerG{3,}{4}{6,}{5}{1,0,0,1-(\alpha_x+j+\zeta/2),-1,1}{0,0.5,-1,-(\alpha_x+j+\zeta/2),1}{bP_{th}L_{rd}\bar{y}}.
	\label{I11closednl}
\end{align}
Then, by substituting \eqref{I11closednl} in \eqref{I1closed}, $\mathbb{I}_1$ is obtained.  
Moreover, $\mathbb{I}_2$ in \eqref{prdNLfintfirst} is given as 
\begin{align}
    \mathbb{I}_2&=\int_{0}^{P_{th}}f_\Phi(\phi)\text{d}\phi =\dfrac{\psi_\phi{P_{th}}^{\alpha_x}}{2}
    \nonumber\\&\times\MeijerG{2,}{2}{2,}{4}{1,1-\alpha_x}{\frac{\xi_x}{2},\frac{-\xi_x}{2},-\alpha_x,1}{\beta_\phi P_{th}}
    \label{I1closi}
\end{align}
where \eqref{I1closi} is calculated by substituting $f_\Phi(\phi)$ from \eqref{fphifi}, and applying \cite[eq. 14]{Adamchik}, \cite[eq. 9.31-1]{Ryzhik43}, and \cite[eq. 26]{Adamchik}.
Then, by substituting \eqref{I1closed}, $\mathbb{P}_{rd}^{\mathcal{D}}(e|\Phi=P_{th})$ from \eqref{prdFconditXint}, and \eqref{I1closi}, SER of the NL-EH model of the link R$\to$D is obtained from \eqref{prdNLfintfirst}.

Finally, the overall upper bound of the BER of the considered DH-DF relaying system under the deterministic distance between the links is obtained by substituting \eqref{Peconditnfixedsr} and \eqref{prdNLfintfirst} in \eqref{PeOverall}.
The following section evaluates in detail the impacts of the system characteristics on the acquired BER expressions.
\section{Performance Evaluation}\label{simResults}
This section presents theoretical and simulation results of the studied cooperative DH-DF relaying in a V2V communications system. 
The results are obtained for various system parameters, which provide a comprehensive insight into the studied system.
The lines and symbols in each figure of this section denote the theoretical and simulation results, respectively.
The simulation and numerical results of the theoretical analysis are in perfect agreement, which validates the mathematical derivations.
The results are provided considering the PS/DA-EH modes as well as the L/NL-EH models.
The results are also obtained for deterministic and uniformly distributed distances of S$\to$R and R$\to$D links.
The integral in \eqref{J1cheyb} is computed using the Gaussian Chebyshev quadrature integral approximation, where the value of $\chi$ in \eqref{J1cheyb} determines the trade-off between complexity and accuracy and is also assumed in \eqref{I11closednl}.
The theoretical BER curves considering PS/DA-EH modes, L/NL-EH models, and deterministic/uniformly distributed distances between S$\to$R/R$\to$D are obtained from \eqref{PeOverall}.

\begin{figure}[t!]
    \centering
    \includegraphics[width=1.1\columnwidth]{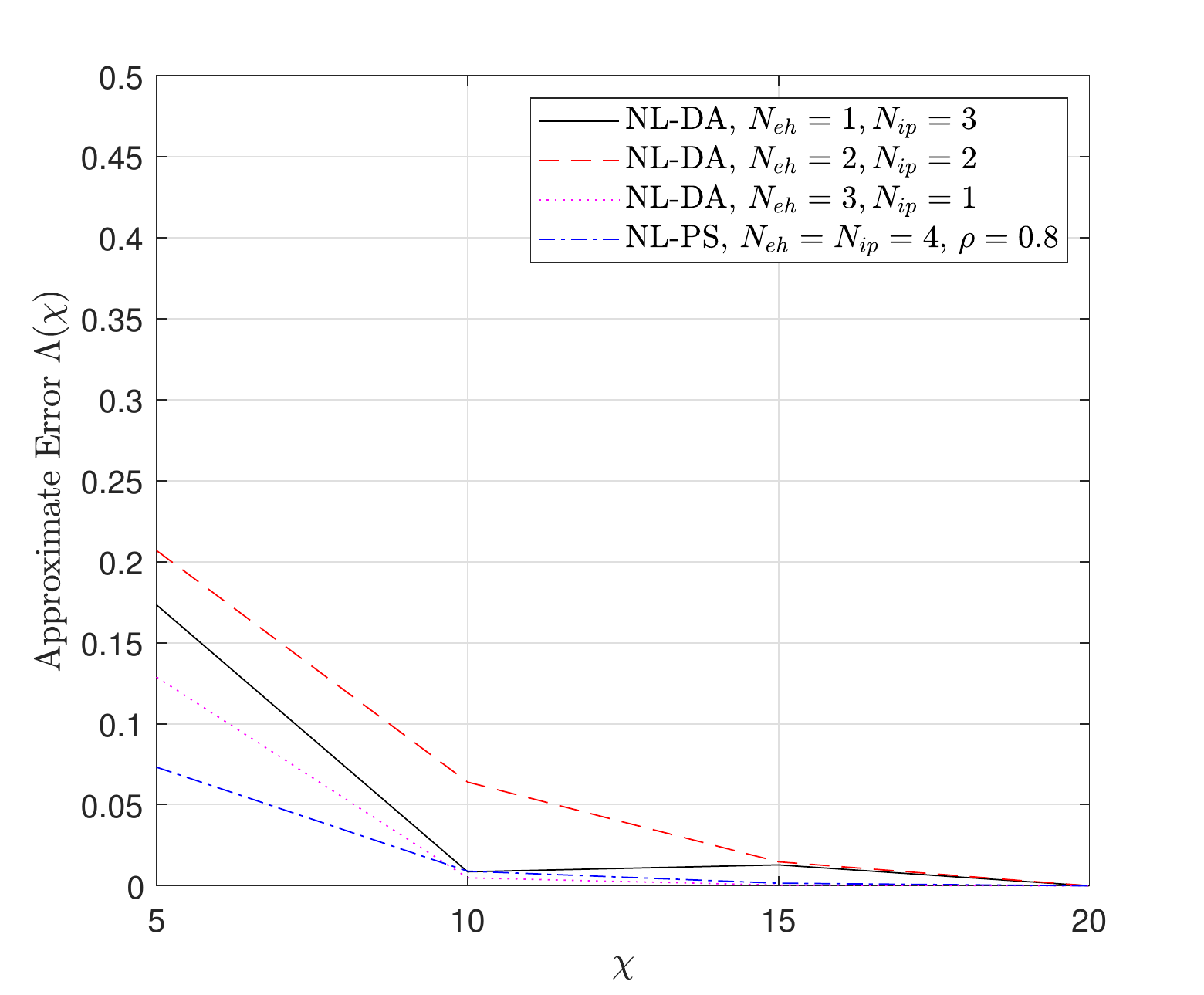}
    \caption{Considered V2V-DH-DF-EH relaying system where $P_t=30$ [dB], $N_{r}=N_{ip}+N_{eh}$, $d_{sr}^U\sim\mathcal{U}(1,3)$, $d_{rd}^U\sim\mathcal{U}(1,3)$.} 
    \label{approxmitErr}
\end{figure}

Unless otherwise stated, we assume that $P_{th}=40$ [dB], $\rho=0.8$ and $\eta=0.7$ \cite{Nasir13,WPT-Reza,piece2wise,PieceWiseModel_MIMO, dipak2022, UAVlNL, Chebishev}.
The modulation order is assumed to be $4$-quadrature amplitude modulation ($4$-QAM), since small data rates are assumed in EH systems.
Moreover, the distances between the S$\to$R and R$\to$D links are uniformly distributed as $d_{sr}^\mathcal{U}\sim\mathcal{U}(1,3)$ and $d_{rd}^\mathcal{U}\sim\mathcal{U}(1,3)$, respectively, and the value of the path-loss exponent is taken as $v=2.7$ \cite{pathlossparameter}.
The channel gains of all links are given as $\Omega_h=\Omega_g=1$, with $\Omega_h=\Omega_{h_i}$ and $n_i=n_{rd}=N_0$ for $i\in\{1,2,\cdots,N_r\}$.
Additionally, node R is equipped with $N_r=4$ receive antennas and one transmit antenna while applying MRC at the receiver side. 
Moreover, $\chi=20$ provides accurate results for both summations given in \eqref{J1cheyb} and \eqref{I11closednl}, where choosing an appropriate $\chi$ is explained in detail in Fig. \ref{approxmitErr}.
\begin{figure}[t!]
    \centering
    \includegraphics[width=1.1\columnwidth]{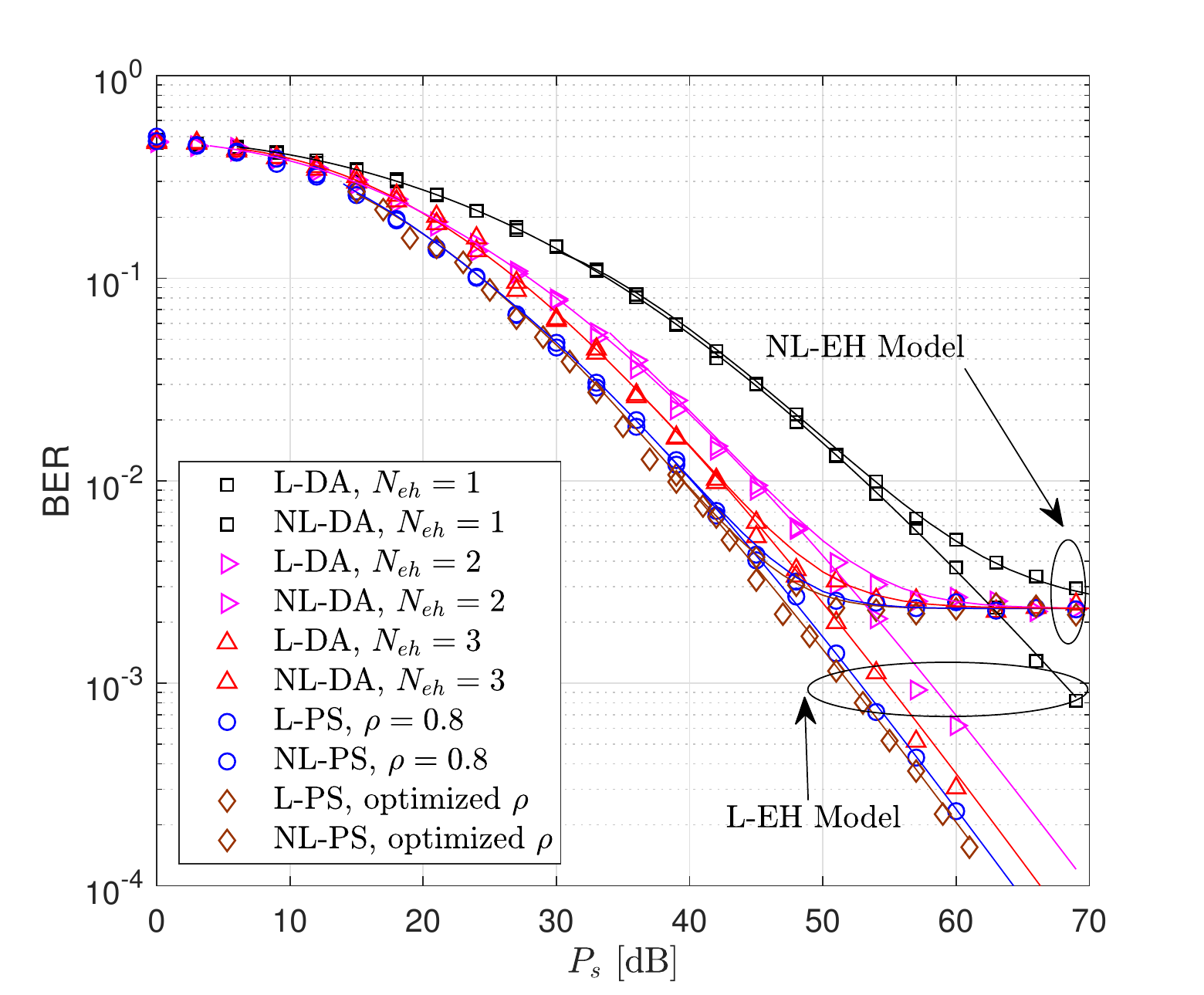}
    \caption{Considered V2V-DH-DF-EH relaying system where $N_r=4$, and $N_{ip}=N_r-N_{eh}$, $d_{sr}^\mathcal{U}\sim\mathcal{U}(1,3)$, $d_{rd}^\mathcal{U}\sim\mathcal{U}(1,3)$.}
    \label{BERps}
\end{figure}

\begin{figure}[t!]
    \centering
    \includegraphics[width=1.1\columnwidth]{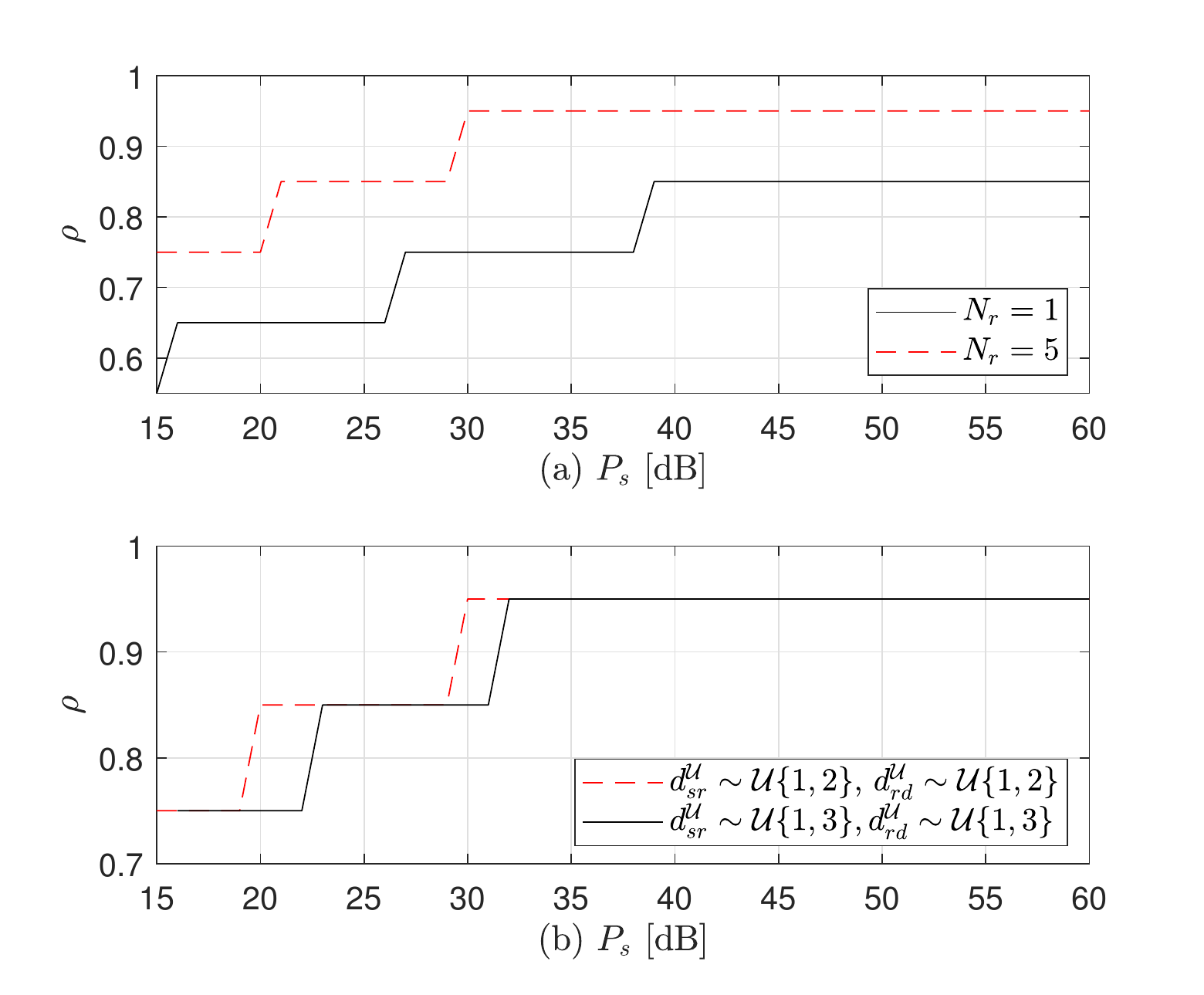}
    \caption{Considered V2V-DH-DF-EH relaying system where (a) $N_r=1$, $N_r=5$
    (b) $d_{sr}^\mathcal{U}\sim\mathcal{U}(1,3)$, $d_{rd}^\mathcal{U}\sim\mathcal{U}(1,3)$, and $d_{sr}^\mathcal{U}\sim\mathcal{U}(1,2)$, $d_{rd}^\mathcal{U}\sim\mathcal{U}(1,2)$.}
    \label{rhoPs}
\end{figure}

The BER performance of a proposed DH-DF relaying system versus $P_s$ [dB] is shown in Fig. \ref{BERps}.
It is assumed that both $d_{sr}^\mathcal{U}$ and $d_{rd}^\mathcal{U}$ are uniformly distributed in the interval $\mathcal{U}(1,3)$ and $\mathcal{U}(1,3)$, respectively, and R is equipped with $N_r=4$ antennas.
The results are given for three different cases of DA-EH mode as $(1,3)$, $(2,2)$ and $(3,1)$ with $(N_{eh},N_{ip})$ and $N_r=N_{eh}+N_{ip}$.
From Fig. \ref{BERps}, it can be seen that the BER performance is increased under the assumption of the DA-EH mode by increasing the number of $N_{eh}$ antennas at node R for both L/NL-EH models.
Moreover, the PS-EH mode with the PS factor $\rho=0.8$ outperforms the three cases of the DA-EH mode with SNR gains of $13$, $4$ and $2$ [dB] for a BER value of $10^{-2}$, respectively, for $N_{eh}=1$, $N_{eh}=2$, and $N_{eh}=3$.
Another result for PS-EH mode in which the $\rho$ value is optimized for each SNR value is given in Fig. \ref{BERps}, while the optimized $\rho$ values are depicted in Fig. \ref{rhoPs}(b). 
This result provides approximately $1$ [dB] SNR gain for SNR values above $P_s=40$ [dB] compared to the PS-EH mode with $\rho=0.8$. However, below SNR values of $P_s=40$ [dB], both L/NL-PS-EH modes provide the same BER performance.
Furthermore, the PS-EH mode provides better performance compared to all three DA-EH modes.
Analysis results are also provided for the NL-EH model in which the BER performance resides in an error floor since the harvested power reaches a saturation power level defined in the energy harvester receiver.
In Fig. \ref{BERps}, it can be seen that the L-EH model overestimates the system performance in high SNR regions compared to the NL-EH model.
Please note that in Fig. \ref{BERps}, numerical results are not given for some SNR values since Meijer's G-function is not defined at these points.

The variation of the $\rho$ value with respect to the SNR values according to PS-EH mode is shown in Fig. \ref{rhoPs}(a) and (b) for $N_r=1,5$, respectively. 
Please note that, for each SNR value, the corresponding $\rho$ value is obtained considering maximum BER performance that is provided in Fig. \ref{rhoPs}(a) and (b).
In Fig. \ref{rhoPs}(a), it is shown that the PS factor $\rho$ is increased directly with an increase in the number of antennas at the R node from $N_r=1$ to $N_r=5$ for both L/NL-EH models, in which at high SNR values the optimized $\rho$ values correspond to $0.85$ and $0.95$ for $N_r=1$ and $N_r=5$, respectively. 
Moreover, the distances of $d_{sr}^\mathcal{U}$ and $d_{rd}^\mathcal{U}$ have an insignificant effect on the $\rho$ value shown in Fig. \ref{rhoPs}(b) and except for $(19,22)$ and $(29,31)$ SNR [dB] ranges, $\rho$ values are approximately equal.

Fig. \ref{approxmitErr} represents the approximate error against the $\chi$ parameter which indicates the trade-off between complexity and accuracy of the calculated integral in \eqref{J1cheyb}. 
We define the approximate error parameter $\Lambda(\chi)$ as the ratio of the difference of simulation value, and the analytical value which is normalized with respect to the simulation value. Here,
the analytical values are calculated from $\mathbb{P}_b^{\mathcal{NL },\mathcal{U}}$ in \eqref{PeOverall} and the simulation results are obtained from Monte-Carlo simulation method considering $P_t=30$ [dB], $N_{r}=N_{ip}+N_{eh}$, $d_{sr}^\mathcal{U}\sim\mathcal{U}(1,3)$, $d_{rd}^\mathcal{U}\sim\mathcal{U}(1,3)$.
From Fig. \ref{approxmitErr}, it can be seen that among all of the obtained results, $\chi=20$ provides a minimum approximation error value equal to $\Lambda\approx0.001$. 
In other words, $\chi=20$ provides an acceptable accurate results for both summations given in \eqref{J1cheyb} and \eqref{I11closednl}.  
\begin{figure}[t!]
    \centering
    \includegraphics[width=1.1\columnwidth]{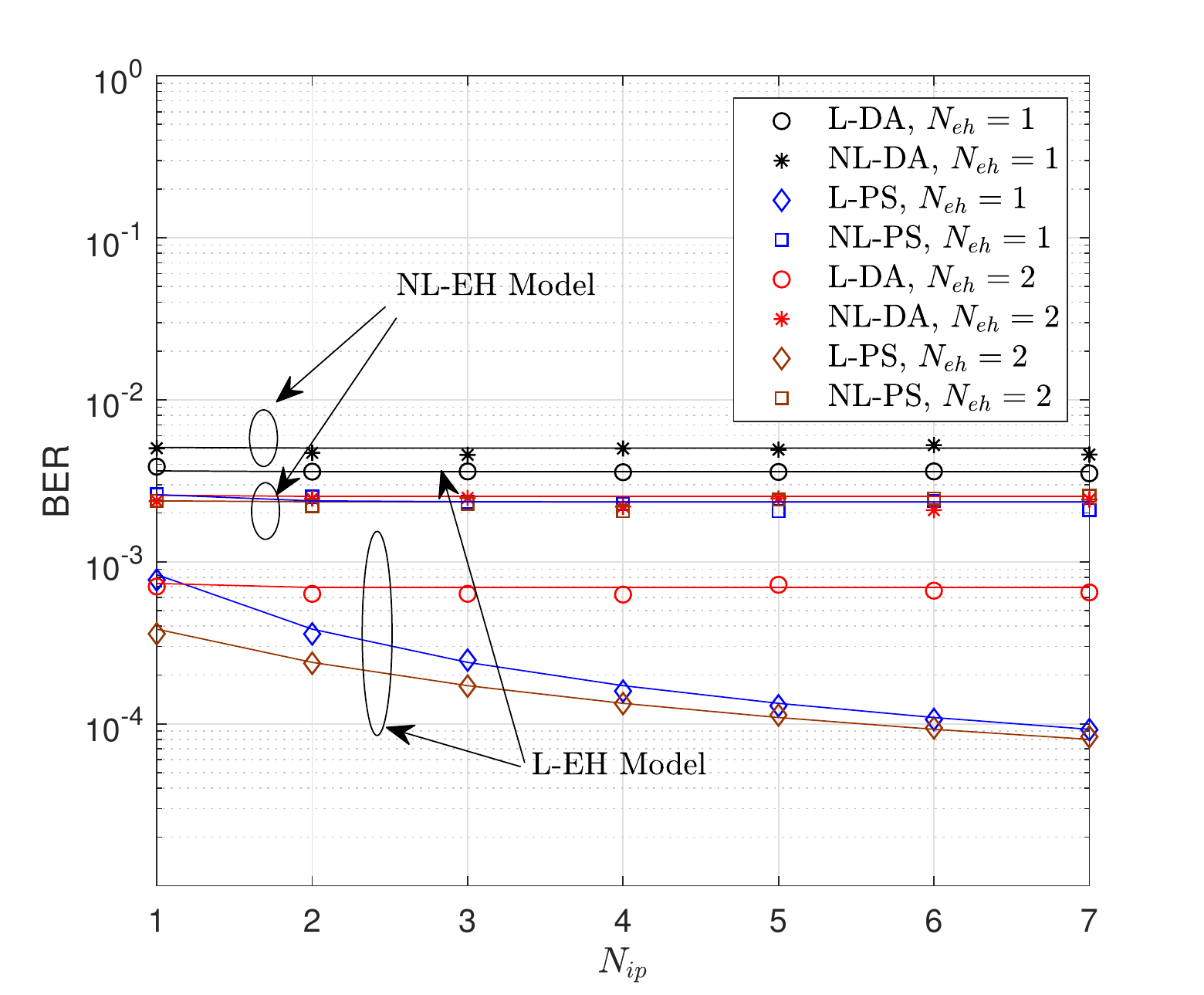}
    \caption{Considered V2V-DH-DF-EH relaying system where $P_t=60$ [dB], $N_{eh}=1,2$, $N_r=N_{ip}+N_{eh}$, and $d_{sr}^\mathcal{U}\sim\mathcal{U}(1,3)$, $d_{rd}^\mathcal{U}\sim\mathcal{U}(1,3)$.}
    \label{Nip1234567}
\end{figure}

The BER performance with respect to $N_{ip}$ is shown in Fig. \ref{Nip1234567}. 
The results are obtained for different numbers of $N_{ip}$ and $N_{eh}=1,2$ antennas, where $N_{r}=N_{ip}+N_{eh}$.
It can be seen that for both $N_{eh}=1$, and $N_{eh}=2$, increasing number of $N_{ip}$ antennas improves the BER performance of the considered system for the L-PS-EH model.
This results from the fact that $N_{r}=N_{ip}+N_{eh}$ number of antennas are concurrently considered for EH and IP. 
For instance, assuming $N_{eh}=1$, and $N_{eh}=2$, R is equipped with $N_{r}=2,\cdots,8$ and $N_{r}=2,\cdots,9$ number of antennas, respectively.      
However, the DA-EH mode uses only $N_{eh}=1$ and $N_{eh}=2$ number of antennas for harvesting energy, which causes low harvested power and low transmit power at the R node and results in a worse received SNR for the R$\to$D link, which effects the overall system performance. 
Therefore, the L-DA-EH model provides an approximate constant BER even when $N_{ip}$ is increased. Moreover, the L-PS-EH model outperforms the L-DA-EH models in all cases. 
Note that, the PS/DA-EH mode provides approximately the same performance considering the NL-EH model for $N_{eh}=2$. This goes to the fact that, for high amounts of harvested power, $P_r=P_{th}$ which results in an error floor and constant BER performance. In other words, the harvested power is saturated to the same predefined $P_{th}$ for both PS/DA-EH modes. 
\begin{figure}[t!]
    \centering
    \includegraphics[width=1.1\columnwidth]{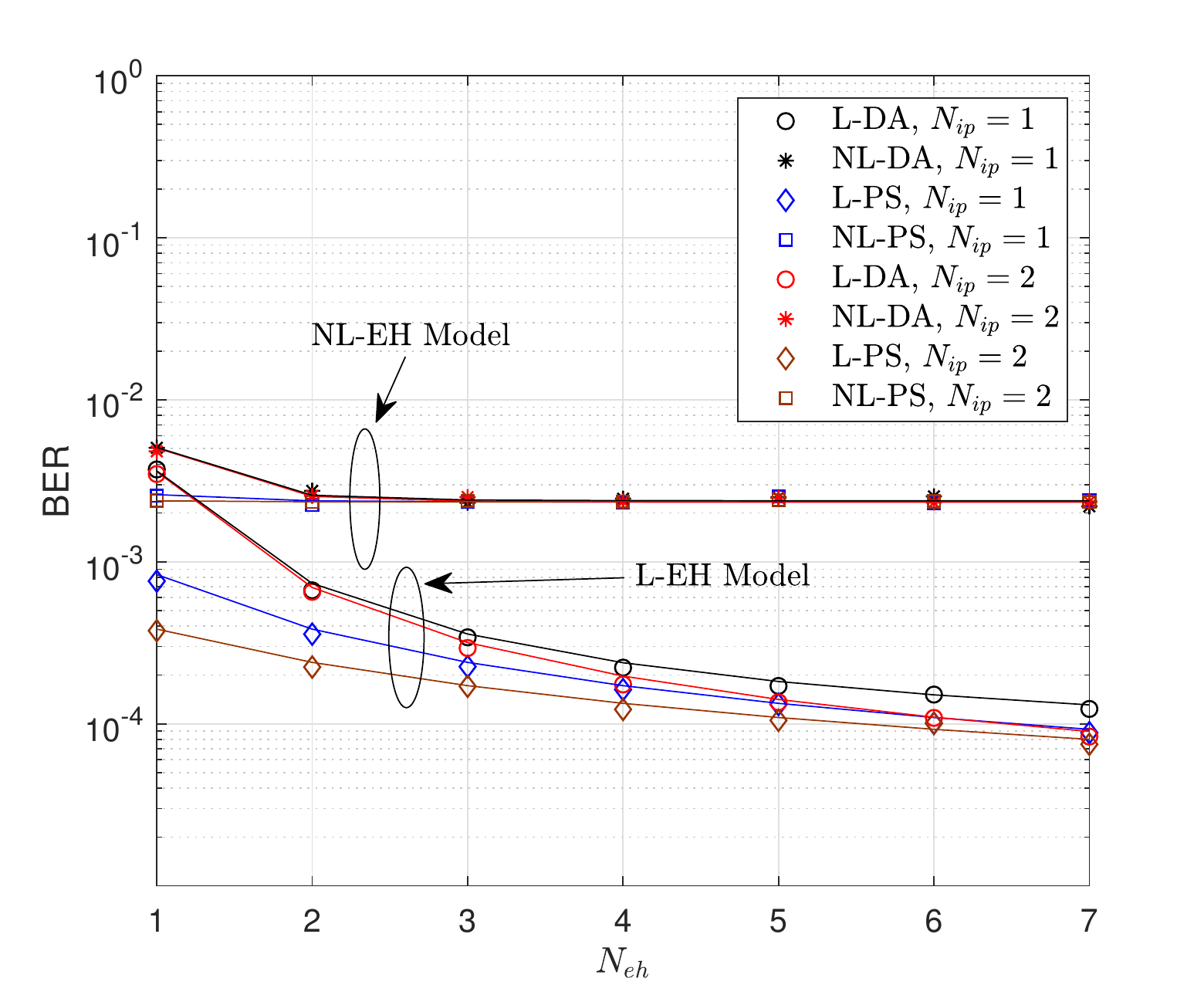}
    \caption{Considered V2V-DH-DF-EH relaying system, where $P_t=60$ [dB], $N_{ip}=1,2$, $N_r=N_{ip}+N_{eh}$, and $d_{sr}^\mathcal{U}\sim\mathcal{U}(1,3)$, $d_{rd}^\mathcal{U}\sim\mathcal{U}(1,3)$.}
    \label{NEH1234567}
\end{figure}

BER performance of the proposed V2V-EH versus $N_{eh}$ is depicted in Fig. \ref{NEH1234567}. 
Apart from Fig. \ref{Nip1234567}, it is shown that the BER performance of L-DA-EH is highly dependent on the number of $N_{eh}$ antennas and the performance improves for both $N_{ip}=1$ and $N_{ip}=2$ as the number of antennas are increased.
However, even with this case, the L-PS-EH model outperforms the L-DA-EH model. Moreover, for all cases of the NL-EH models, the PS/DA-EH modes provide the same performance for both $N_{ip}=1$ and $N_{ip}=2$.  
Finally, considering both Fig. \ref{Nip1234567} and Fig. \ref{NEH1234567}, the BER performance of the NL-EH models are approximately equal.
\begin{figure}[t!]
    \centering
    \includegraphics[width=1.1\columnwidth]{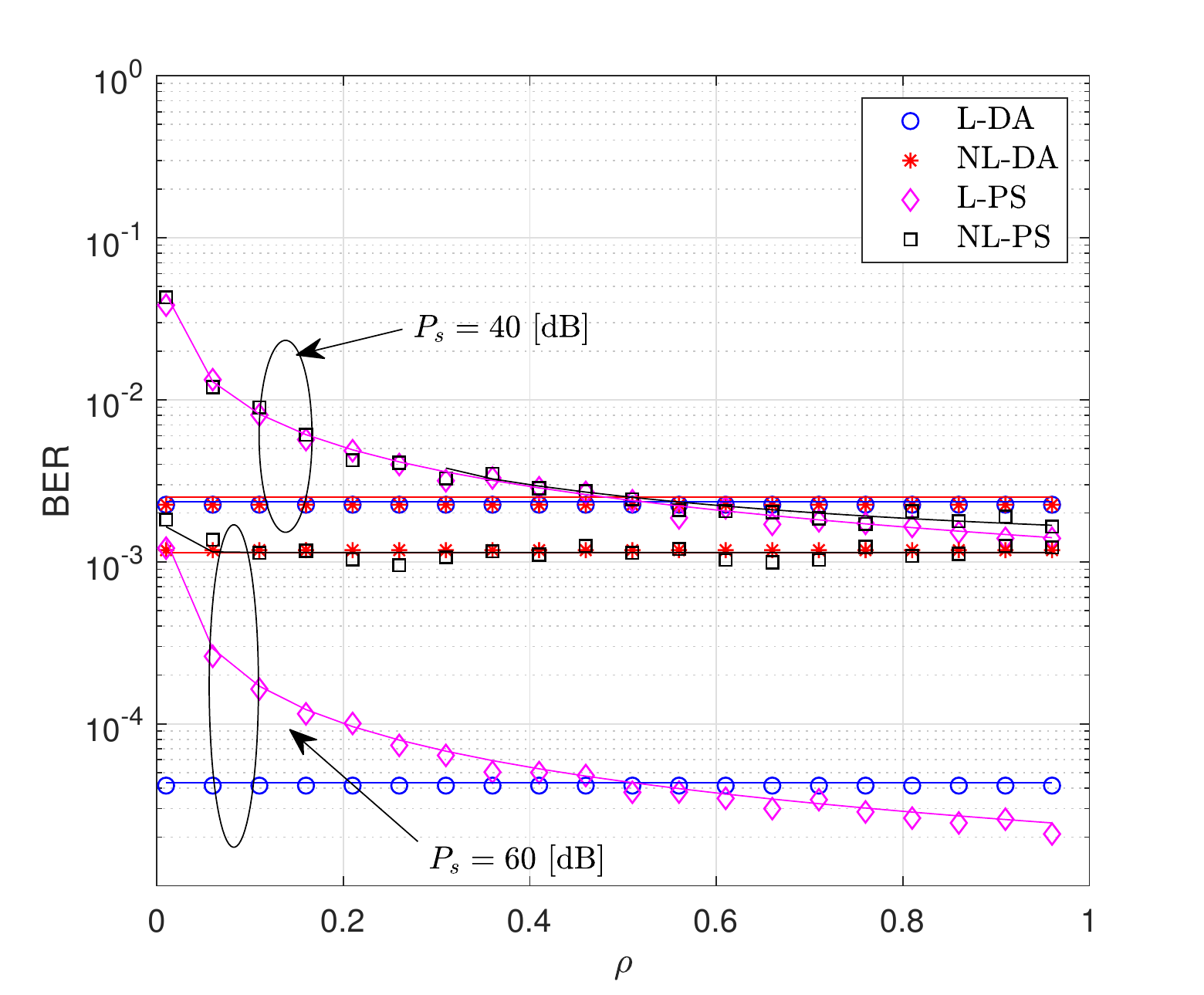}
    \caption{Considered V2V-DH-DF-EH relaying system where $P_t=40,60$ [dB], $N_{eh}=4$, $N_{ip}=2$, $N_r=N_{ip}+N_{eh}$, and $d_{sr}^\mathcal{U}\sim\mathcal{U}(1,2)$, $d_{rd}^\mathcal{U}\sim\mathcal{U}(1,2)$.}
    \label{BERrho40dB60dB}
\end{figure}

The BER of the system with uniformly distributed distances is given in terms of the PS factor $\rho$ in Fig. \ref{BERrho40dB60dB}. The results are obtained for $P_s=40$ [dB] and $P_s=60$ [dB], where $N_{eh}=4$, $N_{ip}=2$, and $d_{sr}^\mathcal{U}\sim\mathcal{U}(1,2)$, $d_{rd}^\mathcal{U}\sim\mathcal{U}(1,2)$.
Note that, since DA is independent of the $\rho$ parameter, the results are the same for all values of $\rho$ and are provided only for comparison purposes. 
From Fig. \ref{BERrho40dB60dB}, it can be seen that the performance of the L-PS-EH model is maximized for a value of $\rho=0.95$ for both $P_s=40$ [dB] and $P_s=60$ [dB]. It is inferred that most of the received signal power at the R node is allocated for harvesting energy at $\rho=0.95$.  
Moreover, the L-DA-EH model provides better error performance compared to the L-PS model for $\rho$ values approximately smaller than $0.5$ for $60$ and $40$ [dB].
Furthermore, the performance of NL-PS/NL-DA-EH is the same for all values of $\rho$ at $P_s=60$ [dB] since R transmit power is reached to the threshold power.
For values larger than $\rho=0.5$, the NL-PS-EH model outperforms the NL-DA-EH model at $P_s=40$ [dB], while for smaller values, the DA-NL EH models provide the optimal BER performance.
\begin{figure}[t!]
	\centering
	\includegraphics[width=1.1\columnwidth]{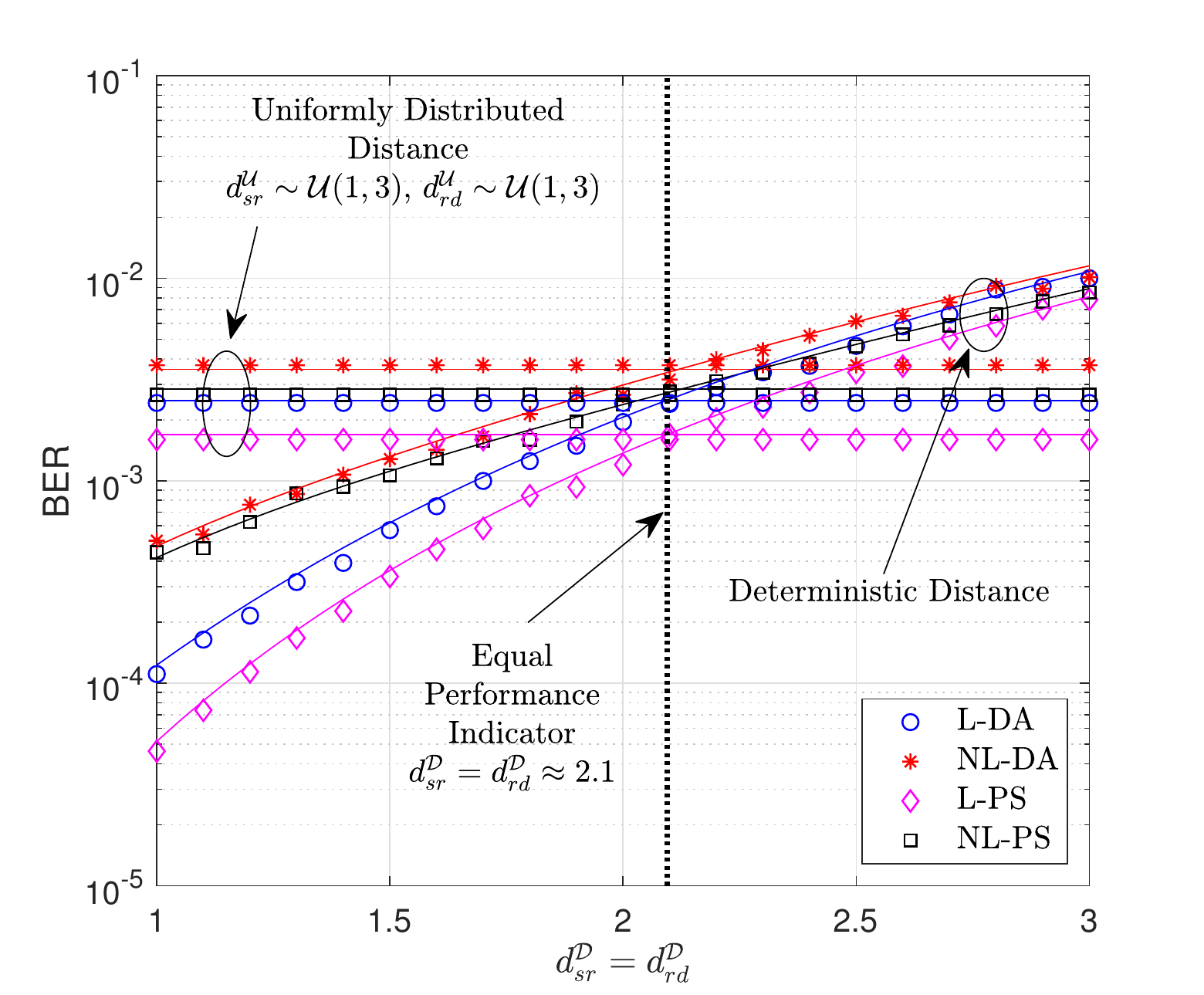}
	\caption{Considered V2V-DH-DF-EH relaying system where $P_t=50$ [dB], $N_{eh}=3$, $N_{ip}=1$, $N_r=N_{ip}+N_{eh}$, and $d_{sr}^\mathcal{U}\sim\mathcal{U}(1,3)$, $d_{rd}^\mathcal{U}\sim\mathcal{U}(1,3)$ for the uniformly distributed distances.}
	\label{BERdsrf13}
\end{figure}

The BER performance of the proposed EH-V2V system is shown in comparison with the deterministic distances $d_{sr}^\mathcal{D}$ and $d_{rd}^\mathcal{D}$ in Fig. \ref{BERdsrf13}. Both distances between S$\to$R and R$\to$D are changed from $1$ to $3$.
It is shown that the system performance deteriorates with increasing distance for both L/NL-EH models of PS/DA-EH modes. 
For a comprehensive analysis, the results are also compared with the case of $d_{sr}^\mathcal{U}\sim\mathcal{U}(1,3)$ and $d_{rd}^\mathcal{U}\sim\mathcal{U}(1,3)$ where both distances are uniformly distributed. 
From Fig. \ref{BERdsrf13}, it can be seen that the performance of the system considering deterministic and random distances is the same for $d_{sr}^\mathcal{D}=d_{rd}^\mathcal{D}=2.1$ for L/NL-EH and PS/DA-EH modes. 
In other words, this means that a system performance with parameters $d_{sr}^\mathcal{U}\sim\mathcal{U}(1,3)$ and $d_{rd}^\mathcal{U}\sim\mathcal{U}(1,3)$ provides the same performance where $d_{sr}^\mathcal{D}=d_{rd}^\mathcal{D}=2.1$.
\begin{figure}[t!]
    \centering
    \includegraphics[width=1.1\columnwidth]{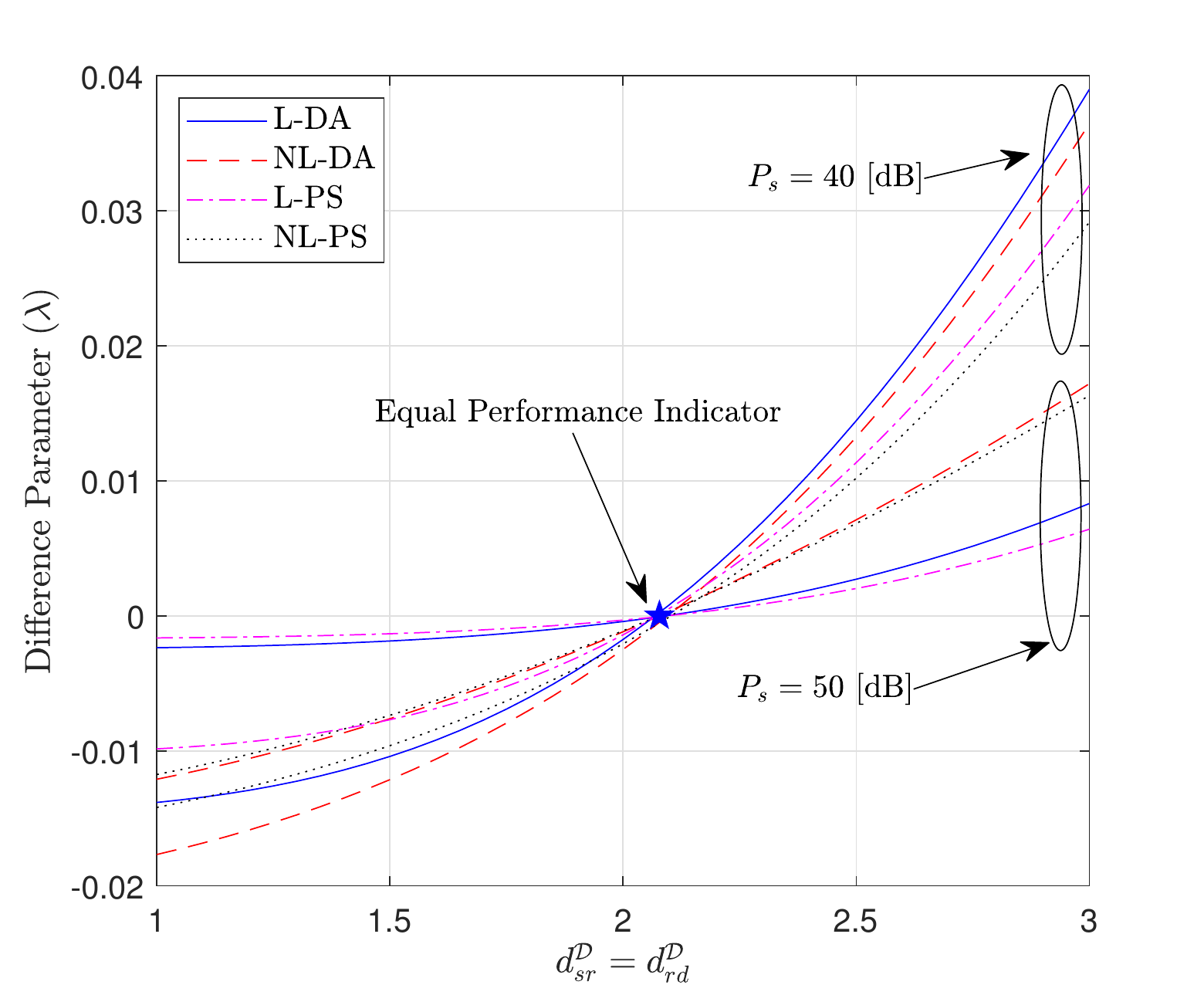}
    \caption{Considered V2V-DH-DF-EH relaying system where $N_{eh}=3$, $N_{ip}=1$, $N_r=N_{ip}+N_{eh}$, and $d_{sr}^\mathcal{U}\sim\mathcal{U}(1,3)$, $d_{rd}^\mathcal{U}\sim\mathcal{U}(1,3)$.}
    \label{differencrlamda}
\end{figure}

Fig. \ref{differencrlamda} represents the difference parameter $\lambda$ versus $d_{sr}^\mathcal{D}=d_{rd}^\mathcal{D}$, where $\lambda$ is calculated from \eqref{PeOverall} as 
\begin{align}
    \lambda=\mathbb{P}^{\mathcal{M},\mathcal{U}}_{b}-\mathbb{P}^{\mathcal{M},\mathcal{D}}_{b}
\end{align} 
and $\mathcal{M}\in\{\mathcal{L},\mathcal{NL}\}$. 
Results are obtained for the PS/DA-EH mode considering both L/NL-EH models for $P_{th}=30$ [dB]. From Fig. \ref{differencrlamda}, it can be seen that both $P_s=40$ and $P_s=50$ [dB] provides equal performance for deterministic and uniformly distributed $d_{sr}^\mathcal{U}\sim\mathcal{U}(1,3)$, $d_{rd}^\mathcal{U}\sim\mathcal{U}(1,3)$ distances at $d_{sr}^\mathcal{D}=d_{rd}^\mathcal{D}=2.1$.
\begin{figure}[t!]
	\centering
	\includegraphics[width=1.1\columnwidth]{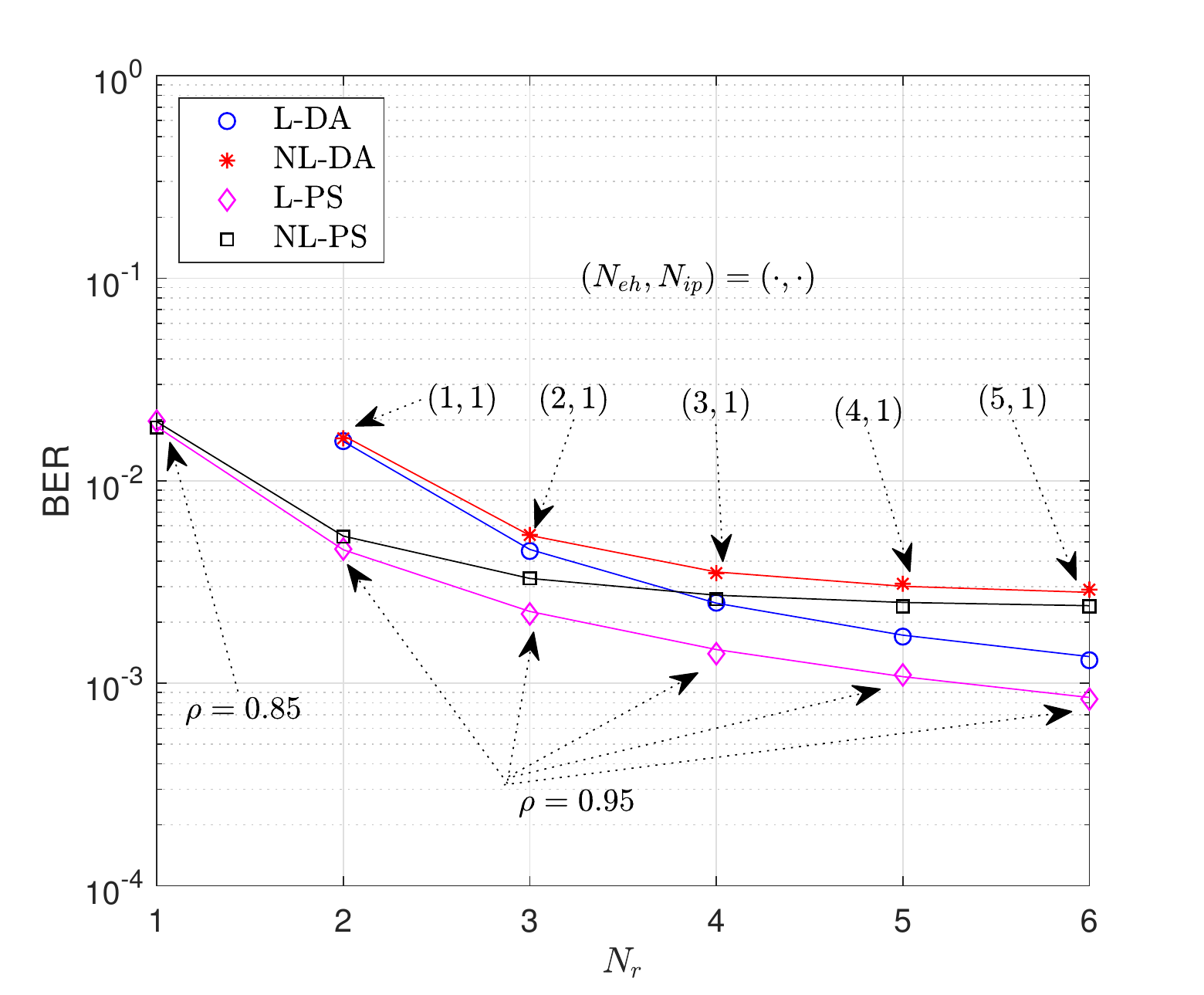}
	\caption{Considered V2V-DH-DF-EH relaying system where $d_{sr}^\mathcal{U}\sim\mathcal{U}(1,3)$, $d_{rd}^\mathcal{U}\sim\mathcal{U}(1,3)$, and $P_t=50$ [dB].}
	\label{BEnroptimiz}
\end{figure}

The BER performance is plotted as a function of $N_r$ number of antennas at the node R in Fig. \ref{BEnroptimiz}. 
The results are provided based on L/NL-EH models and PS/DA-EH modes. 
Each result of the L/NL-EH-PS models is optimized based on $\rho$ value, which are provided in Fig. \ref{BEnroptimiz}. 
It is shown that except for $N_r=1$, $\rho=0.95$ gives the optimum BER performance. This means that high amount of incoming signal power is dedicated to energy harvesting rather than IP. Hence, this increases the transmit power of R.
From Fig. \ref{BEnroptimiz}, it can be seen that $N_{ip}=1$ antenna provides the optimum performance for L/NL-DA-EH models. Therefore, more antennas are dedicated to EH. 
The L-PS-EH model always outperforms the L-DA-EH model.
Furthermore, the same trend is illustrated in NL-PS and NL-DA-EH models.
Note that, when $N_r$ is increased, the performance of NL-PS/NL-DA reaches to the error floor since the harvested power is saturated to $P_{th}$. However, the performance of the L-PS/L-DA-EH model improves with increasing $N_r$, which leads to misinterpretation compared to the practical EH systems.
\section{Conclusion}\label{conclude}
In this paper, the performance of a cooperative vehicular communication system by considering the DF relaying protocol has been investigated jointly with the EH and its application with different EH modes. To become more realistic, distances between the users have been modeled as uniformly distributed random variables and the analyses have been compared with deterministic distance. Besides, in the considered system, the R harvests power from S and employs L/NL-EH models.
Furthermore, the harvested power has been used only for communication purposes in the considered system.
For a comprehensive analysis, the R is assumed to demonstrate the PS/DA-EH modes while applying the MRC scheme to achieve maximum diversity in the system, which is not been extensively studied in the current literature.
The BER performance of the proposed system has been analytically derived and verified by Monte-Carlo simulations considering different system parameters such as distance between nodes, number of antennas at R, and EH modes/models.
It has been shown that PS-EH mode outperforms DA-EH mode under the same conditions, such as using the same number of antennas at R. 
Moreover, the performance of the system under deterministic and uniformly distributed distances has been shown that both performances are approximately converges to each other at the mean value of the uniformly distributed distances. 
It has been perceived that optimal performance has been attained for L/NL-DA-EH models by designating more antennas for EH than IP. In addition, optimal performance has been obtained for L/NL-PS-EH modes by increasing the energy harvesting coefficient.
Furthermore, the L-EH model misrepresented the BER performance at high input powers compared to the practical EH systems, whereas  at low input powers provides acceptable and realistic results.
We investigated the scenario of a single antenna at S and D, however it is feasible to have multiple antennas at S and D and even assess system performance based on transmit antenna selection. However, these assumptions are beyond the scope of our article and will be addressed in a subsequent publication.
\appendices
\section{Pdf of Uniformly Distributed Random Variables}\label{appendixa}
In this part, we assume that the distance is uniformly distributed and the pdf of the path-loss is calculated, accordingly. Then, the path-loss parameter in \eqref{snrSR} is given as 
\begin{align}
    Z=\dfrac{1}{{d_{sr}^\mathcal{U}}^v}
    \label{patheq}
\end{align}
where $v$ is the path-loss exponent. Moreover, $d_{sr}^\mathcal{U}$ is assigned as uniformly distributed as $d_{sr}^\mathcal{U}\sim\mathcal{U}(f,g)$. 
After some mathematical derivations, the pdf of the random variable in  \eqref{patheq} is calculated as
\begin{align}
    f_{Z}(z)=\dfrac{z^{-1-1/v}}{v(g-f)}
    \label{pdfZ}
\end{align}
where $z\in(1/g^v,1/f^v)$. Similarly, the pdf of $W$ in \eqref{snrRD} is calculated as 
\begin{align}
    p_{W}(w)=\dfrac{w^{-1-1/v}}{v(p-r)}
    \label{pdfW}
\end{align}
where $w\in(1/p^v,1/r^v)$ when $d_{rd}^\mathcal{U}\sim\mathcal{U}(r,p)$.
\section{Pdf of $U=XZ$ Random Variables}\label{appendixbfUu}
In this part, the PDF of $f_U(u)$ is calculated as 
\begin{align}
f_U(u)=\int_{g^{-v}}^{f^{-v}}\dfrac{1}{z}f_X(u/z)f_Z(z)\text{d}z
\label{fUuint}
\end{align}
where $f_Z(z)$ and $f_X(x)$ are given in \eqref{pdfZ}, and \eqref{ftehxX}, respectively.
Here, $U=XZ$ denotes the harvested power at R for both DA/PS-EH mode considering \eqref{Harvtspwreanten} and \eqref{HarvtspwrePS}, respectively, under the assumption of uniformly distributed $d_{rd}^{\mathcal{U}}$.
Considering \cite[eq. 14]{Adamchik}, \cite[eq. 9.31-2]{Ryzhik43}, and \cite[eq. 26]{Adamchik}, respectively, and substituting both $f_Z(z)$, and $f_X(x)$ in \eqref{fUuint}, we obtain 
\begin{align}
f_U(u)=\dfrac{\psi_{x}}{2v(g-f)}u^{\alpha_{x}-1}\left(\vartheta(f)-\vartheta(g)\right)
\label{fUufirst}
\end{align}
where 
\begin{align}
\vartheta(\kappa)&={\kappa}^{1+v\alpha_{x}}\nonumber\\&\times\MeijerG{3,}{1}{2,}{4}{1,1-\alpha_{x}-1/v}{\xi_{x}/2,-\xi_{x}/2,-\alpha_{x}-1/v,1}{\beta_{x}{{\kappa}^v}u}
\end{align}
with $\kappa\in\{f,g\}$, respectively.
\ifCLASSOPTIONcaptionsoff
\newpage
\fi
%
\bibliographystyle{IEEEtran}
\bibliography{ref}

\end{document}